\documentclass[fleqn,usenatbib]{mnras}

\usepackage{newtxtext,newtxmath}

\usepackage[T1]{fontenc}

\DeclareRobustCommand{\VAN}[3]{#2}
\let\VANthebibliography\thebibliography
\def\thebibliography{\DeclareRobustCommand{\VAN}[3]{##3}\VANthebibliography}

\usepackage{graphicx}	
\usepackage{amsmath}	
\usepackage{orcidlink}
\usepackage{siunitx}
\usepackage[version=4]{mhchem}
\usepackage[LGRgreek]{mathastext}
\usepackage[normalem]{ulem}

\title[Transport-induced quenching on WASP-96b]{Quenching-driven equatorial depletion and limb asymmetries in hot Jupiter atmospheres: WASP-96b example}

\author[Zamyatina et al.]{Maria Zamyatina$^{{\orcidlink{0000-0002-9705-0535}}1}$\thanks{E-mail: m.zamyatina@exeter.ac.uk},
Duncan A. Christie$^{{\orcidlink{0000-0002-4997-0847}}1,2}$,
Eric H\'ebrard$^{{\orcidlink{0000-0003-0770-7271}}1}$,
Nathan J. Mayne$^{{\orcidlink{0000-0001-6707-4563}}1}$,
\newauthor
Michael Radica$^{{\orcidlink{0000-0002-3328-1203}}3}$,
Jake Taylor$^{{\orcidlink{0000-0003-4844-9838}}3,4}$,
Harry Baskett$^{1}$,
Ben Moore$^{1}$,
Craig Lils$^{1}$,
Denis Sergeev$^{{\orcidlink{0000-0001-8832-5288}}1}$,
\newauthor
Eva-Maria Ahrer$^{{\orcidlink{0000-0003-0973-8426}}2}$,
James Manners$^{5}$,
Krisztian Kohary$^{{\orcidlink{0000-0003-0165-4885}}1}$
and Adina~D.~Feinstein$^{{\orcidlink{0000-0002-9464-8101}}6}$\thanks{NHFP Sagan Fellow}
\\
$^{1}$Department of Physics and Astronomy, Faculty of Environment, Science and Economy, University of Exeter, Exeter EX4 4QL, UK\\
$^{2}$ Max Planck Institute for Astronomy, K\"{o}nigstuhl 17, 69117 Heidelberg, Germany\\
$^{3}$ Institut Trottier de Recherche sur les Exoplanètes and Département de Physique, Université de Montréal, 1375 Avenue Thérèse-Lavoie-Roux,\\ Montréal, QC, H2V 0B3, Canada \\
$^{4}$ Department of Physics (Atmospheric, Oceanic and Planetary Physics), University of Oxford, Parks Rd, Oxford OX1 3PU, UK\\
$^{5}$ Met Office, Fitzroy Road, Exeter EX1 3PB, UK\\
$^{6}$ Laboratory for Atmospheric and Space Physics, University of Colorado Boulder, UCB 600, Boulder, CO 80309
}

\date{Accepted 2024 February 16. Received 2024 February 16; in original form 2023 December 15}

\pubyear{2024}

\begin{document}
\label{firstpage}
\pagerange{\pageref{firstpage}--\pageref{lastpage}}
\maketitle

\begin{abstract}
Transport-induced quenching in hot Jupiter atmospheres is a process that determines the boundary between the part of the atmosphere at chemical equilibrium and the part of the atmosphere at thermochemical (but not photothermochemical) disequilibrium. The location of this boundary, the quench level, depends on the interplay between the dynamical and chemical timescales in the atmosphere, with quenching occurring when these timescales are equal. We explore the sensitivity of the quench level position to an increase in the planet's atmospheric metallicity using aerosol-free 3D GCM simulations of a hot Jupiter WASP-96b. We find that the temperature increase at pressures of $\sim$\SIrange{e4}{e7}{\pascal} that occurs when metallicity is increased could shift the position of the quench level to pressures dominated by the jet, and cause an equatorial depletion of \ce{CH4}, \ce{NH3} and \ce{HCN}. We discuss how such a depletion affects the planet's transmission spectrum, and how the analysis of the evening-morning limb asymmetries, especially within $\sim$\SIrange{3}{5}{\micro\metre}, could help distinguish atmospheres of different metallicities that are at chemical equilibrium from those with the upper layers at thermochemical disequilibrium.
\end{abstract}




\section{Introduction}
\label{sec:introduction}

Hot Jupiters, i.e. close-in extrasolar gas giant planets, were, are and in the near future will be the best targets for exoplanet atmospheric characterisation. These planets host large and often anomalously inflated atmospheres \citep[e.g.,][]{Tremblin2017,Komacek2022}, which aid the detection and identification of atoms and molecules present in these atmospheres. One of the tools used to study hot Jupiter atmospheres is general circulation models \citep[GCMs, e.g.,][]{Showman2009,Rauscher2010,Dobbs-Dixon2013,Mayne2014,Mendonca2016,Sainsbury-Martinez2019,Carone2020,Helling2020,Lee2021,Menou2021}. GCMs simulate the thermodynamic and chemical structure of a planetary atmosphere, and can predict different planetary climates depending on the choice of model parameters. One such parameter is the planet's atmospheric metallicity.

Atmospheric metallicity is the abundance of elements heavier than helium in a planetary atmosphere. Atmospheric metallicity affects many aspects of the planetary climate, e.g., the thermal structure, atmospheric circulation and emitted flux \citep{Drummond2018}, however, the value of atmospheric metallicity for hot Jupiters is uncertain by several orders of magnitude \citep[e.g.,][]{Welbanks2019}. To correctly determine the atmospheric metallicity of a hot Jupiter, one, ideally, needs to fully understand the planet's observed thermochemical state. While achieving such an understanding is an active area of research, hot Jupiters that have ``aerosol-free'' limbs at observable pressures, meaning that the opacities of aerosols constituting condensate clouds and/or photochemical hazes contribute little to the planet's transmission spectrum, can aid such research. WASP-96b was historically considered one of ``aerosol-free'' hot Jupiters. 

WASP-96b is a highly inflated hot Jupiter, with a mass of 0.48$\pm$0.03 $M_J$ and a radius of 1.20$\pm$0.06 $R_J$ \citep{Hellier2014}. It orbits a G8-type star at a distance of 0.0453 AU in 3.425 days, which results in it likely being tidally-locked and having an equilibrium temperature of 1285$\pm$40 K \citep{Hellier2014}. WASP-96b was observed with transit photometry with the Transiting Exoplanet Survey Satellite \citep[\textit{TESS}, \SIrange{0.60}{1.00}{\micro\metre},][]{Yip2021,Nikolov2022} and the Infrared Array Camera on the Spitzer Space Telescope \citep[\textit{Spitzer}/IRAC, 3.6 and \SI{4.5}{\micro\metre},][] {Yip2021,Nikolov2022} as well as with ground- and space-based transit spectroscopy with the FOcal Reducer/low dispersion Spectrograph 2 on the Very Large Telescope Unit 1 \citep[\textit{VLT}/UT1/FORS2, \SIrange{0.35}{0.80}{\micro\metre},][]{Nikolov2018}, the Inamori-Magellan Areal Camera and Spectrograph on the Baade-Magellan Telescope \cite[\textit{Magellan}/IMACS, \SIrange{0.475}{0.825}{\micro\metre},][]{McGruder2022}, the Wide Field Camera 3 using G102 and G141 grisms on the Hubble Space Telescope \citep[\textit{HST}/WFC3 G102 and G141, \SIrange{0.78}{1.65}{\micro\metre},][]{Yip2021,Nikolov2022} and most recently JWST Near Infrared Imager and Slitless Spectrograph using the Single Object Slitless Spectroscopy mode \citep[\textit{JWST}/NIRISS/SOSS, \SIrange{0.6}{2.8}{\micro\metre},][]{Radica2023} as a part of the JWST Early Release Observations (ERO) program \citep{Pontoppidan2022}.

Evidence for the ``aerosol-free'' nature of WASP-96b is three-fold. (1) The WASP-96b \textit{VLT}/UT1/FORS2 transmission spectrum exhibited a broad \ce{Na} I line adsorption feature at $\sim$\SIrange{0.50}{0.75}{\micro\metre} \citep{Nikolov2018}, the presence of which was later independently confirmed by \textit{Magellan}/IMACS observations \citep{McGruder2022}. (2) The WASP-96b \textit{HST}/WFC3 G102 and G141 transmission spectrum showed a full \ce{H2O} absorption feature centered at $\sim$\SI{1.4}{\micro\metre} \citep{Yip2021,Nikolov2022}, the presence of which was also independently confirmed by \textit{JWST}/NIRISS/SOSS observations \citep{Radica2023}. The broadness of the observed \ce{Na} I line feature suggested that the light transmitted through WASP-96b atmosphere was unimpeded by interactions with opacity sources other than \ce{Na}, which would otherwise have obscured \ce{Na} I wings or even its line core. That, in turn, meant that \textit{VLT}/UT1/FORS2 and \textit{Magellan}/IMACS observations probed a large range of the planet's atmospheric pressures and captured \ce{Na} I adsorption line broadening \citep{Hedges2016}. Similarly to \ce{Na} I, the observation of a full and unmuted \ce{H2O} absorption feature suggested that \ce{H2O} was the only opacity source contributing to the spectrum at \textit{HST}/WFC3 G102 and G141 and relevant \textit{JWST}/NIRISS/SOSS wavelengths. (3) 1D forward and retrieval modelling based on the aforementioned WASP-96b pre-JWST transmission spectrum had consistently found WASP-96b's observed limbs to be ``aerosol-free'' \citep{Nikolov2018,Welbanks2019,Alam2021,Nikolov2022,McGruder2022}.

The ``aerosol-free'' status of WASP-96b has recently been put into question by the theoretical work of \cite{Samra2023} and observations made using \textit{JWST}/NIRISS/SOSS \citep{Radica2023,Taylor2023}. \cite{Samra2023} explored the possibility of cloud formation in the atmosphere of WASP-96b using a hierarchy of forward and retrieval models. Their simulations predict that a mix of $\approx$40\% silicate (\ce{MgSiO3}[s], \ce{Mg2SiO4}[s], \ce{Fe2SiO4}[s], \ce{CaSiO3}[s]) and $\approx$40\% metal oxide (\ce{SiO}[s], \ce{SiO2}[s], \ce{MgO}[s], \ce{FeO}[s], \ce{Fe2O3}[s]) clouds would dominate the limbs of WASP-96b at observable pressures and that there would be a slight asymmetry between the limbs, with the morning limb being cloudier. They proposed two scenarios that allow for \ce{Na} I and \ce{H2O} features to be observed even in the presence of clouds. (1) WASP-96b could have a low vertical mixing efficiency, which would reduce the cloud top height whilst keeping the cloud base unchanged. (2) WASP-96b could have a high cloud particle porosity, which would reduce the cloud opacity by making the clouds optically thinner. \citet{Radica2023} and \citet{Taylor2023} reported tentative evidence for haze in the atmosphere of WASP-96b. According to \citet{Radica2023}, the first \textit{JWST} spectrum of WASP-96b contained three \ce{H2O} features centered at $\sim$1.2, 1.4 and \SI{1.8}{\micro\metre}, respectively, and a \ce{K} feature at $\sim$\SI{0.76}{\micro\metre} embedded into a Rayleigh scattering slope. Their analysis of this spectrum with three grids of 1D radiative-convective chemical equilibrium models revealed the following. (1) \ce{MnS}[s] and \ce{MgSiO3}[s] cloud formation, inferred from comparing their best-fitting planet-average solar\footnote{With solar abundances from \citet{Lodders2009}.}-metallicity pressure-temperature profile to several cloud species' condensation curves, is possible, however the cloud top of such, assumed to be optically thick, clouds had to be located below the pressures observable during transit (at $\sim$\num{e5} \unit{\pascal}). (2) The observed Rayleigh slope could be explained by either an enhanced Rayleigh scattering by small particles, i.e., possibly haze, or by a highly-pressure-broadened red wing of the \ce{Na} I line. The degeneracy between an enhanced Rayleigh scattering and a high \ce{Na} abundance had been pointed out previously \citep[e.g., by][]{Nikolov2018}, but breaking this degeneracy in the case of WASP-96b would require an analysis of, e.g., a WASP-96b combined \textit{VLT}/UT1/FORS2 and \textit{JWST}/NIRISS/SOSS spectrum, which was outside the scope of their study. 
\citet{Taylor2023} extended the work of \citet{Radica2023} by interpreting WASP-96b \textit{JWST}/NIRISS/SOSS observations using a combination of two 3D GCMs and four grids of 1D retrievals. Both participating GCMs, the \textsc{SPARC/MITgcm} and the Met Office \textsc{Unified Model} (\textsc{UM}), were aerosol-free and were run for two scenarios: the \textsc{SPARC/MITgcm} with 1$\times$ and 10$\times$solar\footnote{With solar abundances from \citet{Lodders2002a}.} metallicity under the assumption of chemical equilibrium, and the UM with 1$\times$solar\footnote{With solar abundances from \citet{Asplund2009} and \citet{Caffau2011}.} metallicity and under the assumption of chemical equilibrium or disequilibrium. The comparison of GCM-derived limb-average pressure-temperature profiles to cloud species' condensation curves revealed that all four GCM simulations predicted a thermal structure that permits \ce{MnS}[s] cloud formation at both WASP-96b's limbs, with \textsc{SPARC/MITgcm} 1$\times$solar simulation also allowing for high-altitude \ce{Na2S}[s] clouds at the morning limb. \ce{MgSiO3}[s] cloud formation was also possible at $\geq$\num{e5} \unit{\pascal}. The transmission spectra computed from GCM simulations generally agreed with the \textit{JWST}/NIRISS/SOSS spectrum longward of \SI{1.3}{\micro\metre}, but underestimated the observed transit depths shortward of \SI{1.3}{\micro\metre} likely due to the lack of a scattering opacity source. That encouraged an investigation using atmospheric retrievals. All retrieval codes participating in the work of \citet{Taylor2023}, \textsc{CHIMERA}, \textsc{Aurora}, \textsc{PyratBay} and \textsc{POSEIDON}, found that scenarios with inhomogeneous clouds and hazes were preferred over cloud-free ones, which was caused primarily by the detection of a Rayleigh scattering slope rather than of an opacity from a grey cloud top. All free retrieval codes recovered chemical species abundances that are consistent between the codes, with abundances from \textsc{Aurora} being: $\log_{10}(\ce{H2O})=-3.59${\raisebox{0.5ex}{\tiny$\substack{+0.35 \\ -0.35}$}}, $\log_{10}(\ce{CO2})=-4.38${\raisebox{0.5ex}{\tiny$\substack{+0.47 \\ -0.57}$}}, $\log_{10}(\ce{K})=-8.04${\raisebox{0.5ex}{\tiny$\substack{+1.22 \\ -1.71}$}}, which are generally consistent with a solar\footnote{With solar abundances from \citet{Lodders2002a}.} metallicity and a solar \ce{CO} (upper limit), except for \ce{CO2}, whose abundance is more consistent with its equilibrium abundance at 10$\times$solar metallicity.

In addition to aerosol formation, other processes can alter the transmission spectrum of hot Jupiters atmospheres, e.g., transport-induced quenching \citep{Moses2014,Zahnle2014}, which we later refer to as ``quenching'' for simplicity. Quenching is a process that determines the boundary between the part of the atmosphere at chemical equilibrium and the part of the atmosphere at thermochemical (but not photothermochemical) disequilibrium. The location of this boundary, the quench level, depends on the interplay between the dynamical and chemical timescales in the atmosphere, with quenching occurring when these timescales are equal. The sensitivity of quenching to changes in a variety of stellar and planetary parameters was explored with pseudo-2D \citep{Baeyens2021,Moses2022} and 3D \citep{Drummond2020,Zamyatina2023} forward models, however, the sensitivity of quenching to the metallicity increase has never been explored with a GCM.

Motivated by the uncertainty in WASP-96b's atmospheric metallicity, in this study we designed several 3D GCM simulations (Section~\ref{sec:simulations}) to help us do the following. (1) Explore how 3D GCM predictions of WASP-96b's atmospheric circulation (Section~\ref{sec:atmospheric_circulation}) and chemistry (Section~\ref{sec:chemical_structure}) differ depending on the assumed metallicity and the assumed atmospheric thermochemical state, equilibrium or disequilibrium. (2) Explore how quenching responds to the metallicity increase, and causes an equatorial depletion of \ce{CH4}, \ce{NH3} and \ce{HCN} (Section~\ref{sec:equatorial_depletion}) and limb asymmetries (Section~\ref{sec:limb_asymmetries}). (3) Identify parts of the simulated WASP-96b transmission spectrum that are sensitive to the metallicity increase and disequilibrium chemistry (Section~\ref{sec:sensitivity_to_mdh_and_quenching}). Finally, we compare the observed and simulated WASP-96b transmission spectra (Section~\ref{sec:comparison_to_existing_obs}), and present our discussion and conclusions in Sections~\ref{sec:discussion} and \ref{sec:conclusions}, respectively.

\section{Simulations}
\label{sec:simulations}

We used the Met Office \textsc{Unified Model} (\textsc{UM}) GCM with the same basic model setup as in \citet{Drummond2020} and \citet{Zamyatina2023}. In brief, the dynamical core of our GCM solves the full, deep-atmosphere, non-hydrostatic equations of motion using a semi-implicit semi-Lagrangian scheme. The radiative transfer component of our GCM solves the two-stream equations in 32 spectral bands covering \SIrange{0.2}{322}{\micro\metre} and treats opacities using the correlated-k and equivalent extinction methods. Our opacity sources include the absorption due to \ce{H2O}, \ce{CO}, \ce{CO2}, \ce{CH4}, \ce{NH3}, \ce{HCN}, \ce{Li}, \ce{Na}, \ce{K}, \ce{Rb}, \ce{Cs} and collision-induced absorption due to \ce{H2}-\ce{H2} and \ce{H2}-\ce{He} as well as Rayleigh scattering due to \ce{H2} and \ce{He} (see \citet{Goyal2020}, for line list information). We chose a model grid resolution of \ang{2.5} longitude by \ang{2} latitude and 66 vertical levels equally spaced in height (covering pressures from $\sim$\SI{2e7}{\pascal} to $\sim$\SI{1}{\pascal}). While such a grid resolution is too coarse to resolve individual convective plumes, we let the non-hydrostatic core of our GCM remove convective instabilities at the grid scale and do not use a convection scheme. For the stellar spectrum we used the \textsc{PHOENIX} BT-Settl stellar spectrum \citep{Rajpurohit2013} closely matching WASP-96 (Table~\ref{tab:star_parameters}), and for the planet --- WASP-96b parameters from \citet{Hellier2014} (Table~\ref{tab:planet_parameters}). We initialised our GCM simulations at rest with the dayside-average pressure-temperature profiles from the 1D radiative-convective-chemistry model \textsc{ATMO} \citep[e.g.,][]{Tremblin2015,Drummond2016} adopting the same WASP-96 system parameters and assuming chemical equilibrium for the chemical species present in the \citet{Venot2012} chemical network. The intrinsic temperature at the bottom boundary of the model domain was assumed to be \SI{100}{\kelvin} in both ATMO and the UM. Lastly, and contrary to the GCM simulations in \citet{Drummond2020}, \citet{Zamyatina2023} and \citet{Taylor2023}, we used a smaller filtering constant, $K$, set to $0.04$, instead of previously used $K=0.16$, to reduce the strength of longitudinal filtering of the horizontal wind \citep[e.g.,][]{Mayne2017}, i.e., effectively reducing diffusion, to prevent the loss of a considerable amount of axial angular momentum and the emergence of a retrograde equatorial jet, both discussed in detail in Christie et al. (in prep.). All GCM simulations were aerosol-free, and were run for at least 1000 Earth days to let the upper atmosphere (\num{e2}-\num{e5} \unit{\pascal}) reach a pseudo-steady state dynamically, radiatively and chemically (Figs.~\ref{fig:wasp96b_conservation}-\ref{fig:wasp96b_steady_state}). All fields shown later in this paper were chosen to be averaged over the last 200 simulation days.

\begin{table}
\caption{WASP-96 parameters used in this study.}
\label{tab:star_parameters}
\centering
\begin{tabular}{lcc}
\hline
Parameter & Value & Unit\\
\hline
Type & G8 & \\
Radius & \num{7.30e+08} & \unit{\metre}\\ 
Effective temperature & 5500 & \unit{\kelvin}\\
Stellar constant at 1 au & 1272.86 & \unit{\watt\per\square\metre}\\ 
$\log_{10}$(surface gravity) & 4.50 & (cgs)\\
Metallicity & 0.00 & dex\\
\hline
\end{tabular}
\end{table}

\begin{table}
\centering
\caption{WASP-96b parameters used in this study. Two values are listed for parameters that change with changing metallicity.}
\label{tab:planet_parameters}
\begin{tabular}[t]{lcc}
\hline
Parameter & Value & Unit\\
          & (for 1$\times$, 10$\times$solar)  &\\                         
\hline
Inner radius & \num{8.39e+07} & \unit{\metre}\\ 
Domain height & \num{1.03e+07} & \unit{\metre}\\ 
Semi-major axis & 0.0453 & au \\
Orbital period & 3.4252602 & Earth day\\
Rotation rate& \num{2.12e-05} & \unit{\radian\per\second}\\ 
Surface gravity & 7.59 & \unit{\metre\per\square\second}\\ 
Specific gas constant& 3516.29, 3164.69 & \unit{\joule\per\kelvin\per\kilogram}\\ 
Specific heat capacity& \num{1.26e+04}, \num{1.15e+04} & \unit{\joule\per\kelvin\per\kilogram}\\ 
Stellar irradiance & \num{0.62e+06} & \unit{\watt\per\square\metre}\\ 
Effective temperature & $1250$ & \unit{\kelvin}\\
\hline
\end{tabular}
\end{table}

We performed two pairs of GCM simulations (i.e., four simulations in total), exploring the sensitivity of the atmospheric structure of WASP-96b to the following: 
\begin{itemize}
    \item[(1)] the assumed metallicity, 1$\times$ solar ($[M/H]=0$) or 10$\times$ solar ($[M/H]=1$); and
    \item[(2)] the assumption about whether the entire atmosphere is at chemical equilibrium or allowing upper layers of the atmosphere to be at chemical disequilibrium. The simulations that assumed chemical equilibrium computed abundances of chemical species using the Gibbs minimisation scheme, and are later referred to as ``equilibrium'' simulations. The simulations that allowed for chemical disequilibrium computed the production and loss of chemical species present in the \citet{Venot2019} reduced chemical network, and are later referred to as ``kinetics'' simulations. 
\end{itemize}

We did not include photochemistry into our kinetics simulations, meaning that these simulations take into account only disequilibrium thermochemistry. We also assumed gas-phase only composition in all simulations. Finally, our simulations included the calculation of alkali metal abundances during the simulation run time. Since the chemical kinetics of alkali metals is poorly known for \ce{H2}-rich atmospheres, alkali metals were not included into the chemical network used in our kinetics simulations. Instead, in both equilibrium and kinetics simulations, alkali metal abundances were computed using the parameterization of \citet{Amundsen2016}, which takes alkali metals’ monatomic/polyatomic transformation curves from \citet{Burrows1999} chemical equilibrium abundances and applies additional smoothing to these curves to avoid a step change in abundance. We plan to transition from using this parameterization to computing alkali metal abundances assuming chemical equilibrium in the near future.

\section{Results}
\label{sec:results}


\subsection{Atmospheric circulation}
\label{sec:atmospheric_circulation}

\begin{figure*}
\includegraphics[scale=0.64]{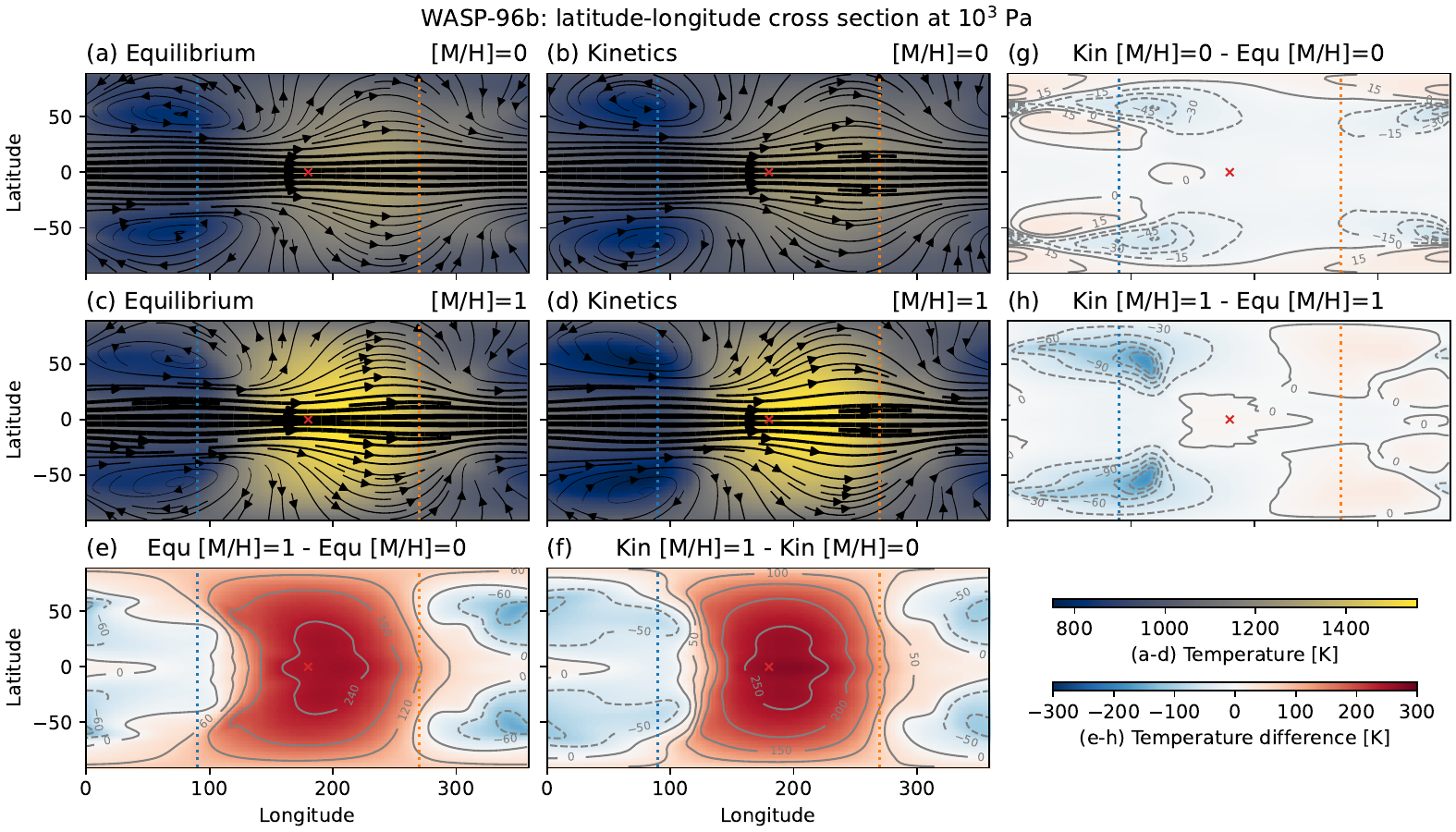}
 \caption{Temperature and horizontal wind streamlines at \num{e3} \unit{\pascal} in the WASP-96b equilibrium and kinetics simulations assuming $[M/H]=0$ or $[M/H]=1$. Panels (a-d) show the general circulation pattern: the equatorial jet and mid-latitude Rossby wave crests and troughs. Panels (a-d) share a colour scale (the top one in the bottom right of the figure), and a streamline scale, with the width of streamlines scaled with horizontal wind speed and normalised to the maximum horizontal wind speed within \num{e2}-\num{e5} \unit{\pascal} (i.e., effectively inside the equatorial jet) between all four simulations. Panels (e-f) show the difference caused by an increase in metallicity, and (g-h) the difference caused by disequilibrium thermochemistry. Panels (e-h) share a colour scale (the bottom one in the bottom right of the figure). Red crosses mark the substellar point, and vertical dotted lines mark the morning (\ang{90}E in blue) and evening (\ang{270}E in orange) limb.}
 \label{fig:wasp96b_hwind_temp_at_plev}
\end{figure*}

Let us consider the atmospheric circulation and temperature of WASP-96b predicted at \SI{e3}{\pascal} (Fig.~\ref{fig:wasp96b_hwind_temp_at_plev}a-d). As expected for any typical tidally-locked hot Jupiter, the dayside of the planet is hotter than the nightside, and the hot spot is shifted eastward from the substellar point by the zonal heat transport within the eastward equatorial jet. Both the metallicity increase (Fig.~\ref{fig:wasp96b_hwind_temp_at_plev}e-f) and disequilibrium thermochemistry (Fig.~\ref{fig:wasp96b_hwind_temp_at_plev}g-h) have an impact on the planet's thermodynamical structure, but an increase in metallicity alters this structure more than disequilibrium thermochemistry.

\begin{figure*}
\includegraphics[scale=0.64]{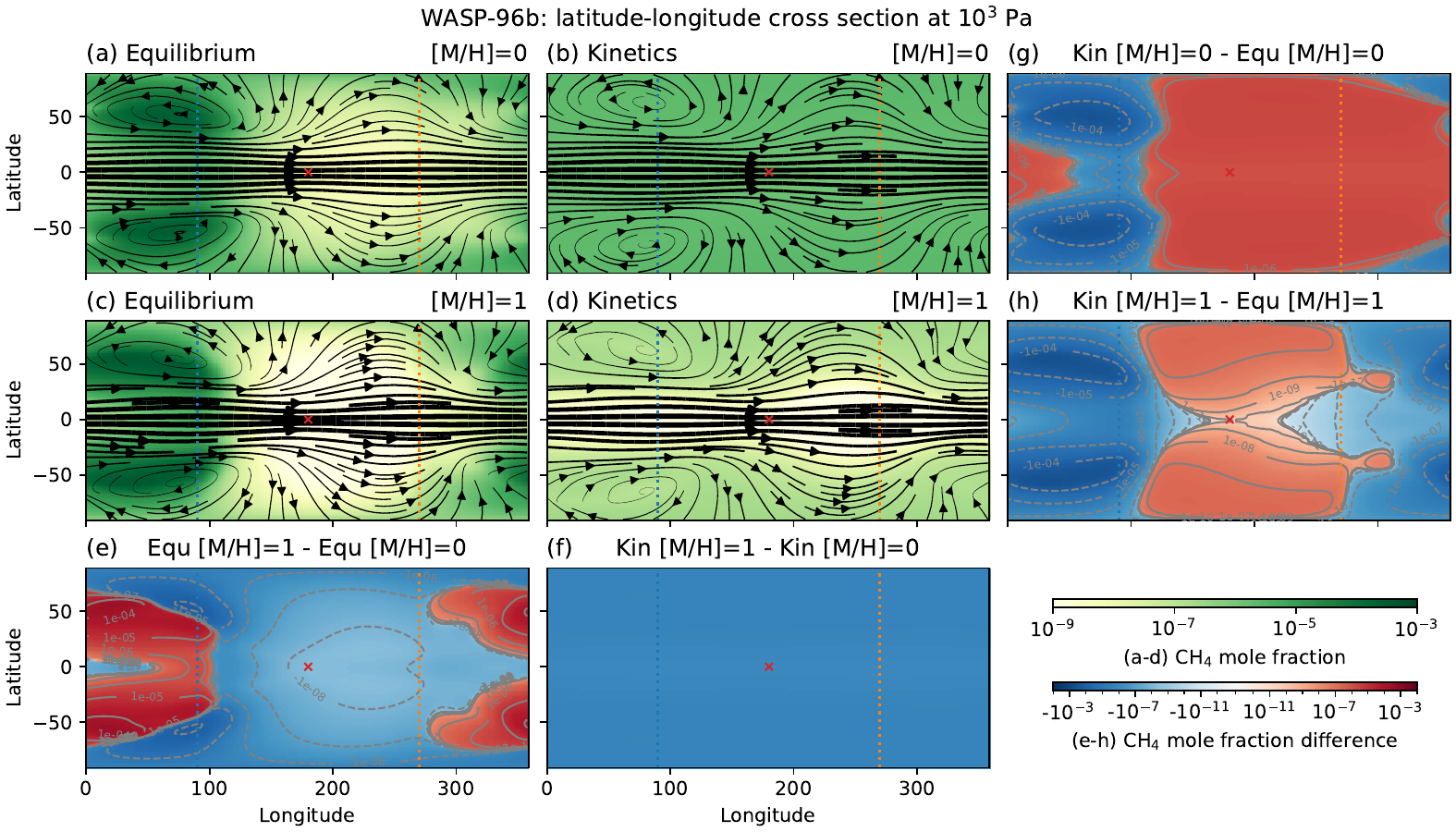}
 \caption{As in Fig.~\ref{fig:wasp96b_hwind_temp_at_plev} but for \ce{CH4} mole fraction.}
 \label{fig:wasp96b_hwind_ch4_at_plev}
\end{figure*}

An increase in metallicity from $[M/H]=0$ to $[M/H]=1$ increases the day-night temperature contrast at \SI{e3}{\pascal}. At $[M/H]=0$ the day-night temperature contrast (dayside maximum minus nightside minimum) is $\sim$\SI{500}{\kelvin}, while at $[M/H]=1$ it is higher, mostly due to a higher dayside temperature, and is equal to $\sim$\SI{800}{\kelvin}. This temperature contrast is accompanied by the difference in the maximum zonal-mean zonal wind speed (Fig.~\ref{fig:wasp96b_pres_u_znl_mean}). This wind speed is $\sim$\SI{0.5}{\km\per\second} faster at $[M/H]=1$ than at $[M/H]=0$, whilst the jet occupies a similar part of the atmosphere --- between $\pm$\ang{40} latitude and down to $\sim$\num{3e5} \unit{\pascal}. Such a change in WASP-96b's thermodynamical structure with increasing metallicity occurs because as metallicity increases the mean molecular weight and opacity of the atmosphere increases while its specific heat capacity decreases \citep[e.g.,][]{Drummond2018}. The combined effect of these changes causes the atmosphere to experience a more compact flow structure and shallower heating, which together reduce the day-night heat transport, and cause the simulations with a higher metallicity to have an overall warmer dayside and a cooler nightside than the simulations with a lower metallicity.

An increase in metallicity from $[M/H]=0$ to $[M/H]=1$ also amplifies the thermal limb\footnote{Limb here is defined as a pressure-latitude cross-section that divides a planetary atmosphere into a day-lit side and a dark night side. In our simulations, this cross-section aligns with longitudes \ang{90}E and \ang{270}E, with the substellar point being at \ang{180}E.} asymmetry at \SI{e3}{\pascal}. Given that a tidally-locked hot Jupiter typically has a colder morning limb and a warmer evening limb, our simulations predict that at \SI{e3}{\pascal} WASP-96b's cold morning limb will be even colder (by up to \SI{50}{\kelvin}) and the warm evening limb will be even warmer (by up to \SI{100}{\kelvin}) at $[M/H]=1$ than at $[M/H]=0$. This results in the average evening minus morning limb temperature difference at \SI{e3}{\pascal} being $\sim$\SI{250}{\kelvin} at $[M/H]=0$ and $\sim$\SI{460}{\kelvin} at $[M/H]=1$. The thermal limb asymmetry at other pressure levels is shown in Fig.~\ref{fig:wasp96b_vp_pres_temp_limbs}.

Disequilibrium thermochemistry changes the predicted thermal structure of WASP-96b at \SI{e3}{\pascal} mostly near the morning limb. The morning limb is influenced by the circulation in the Rossby gyres \citep[e.g.,][]{Showman2020} that transport colder gas from the nightside to the dayside in the equatorial region and warmer gas from the dayside to the nightside in the polar regions. At both metallicities our kinetics simulations predict lower temperatures inside the Rossby gyres than those in the equilibrium simulations. As a result, the morning limb in the kinetics simulations is colder up to \SI{30}{\kelvin} at $[M/H]=0$ and \SI{120}{\kelvin} at $[M/H]=1$ (i.e. comparable in magnitude to the thermal limb asymmetry due to the metallicity increase).

\subsection{Chemical structure}
\label{sec:chemical_structure}

The predicted chemical structure of WASP-96b changes considerably due to both the metallicity increase from $[M/H]=0$ to $[M/H]=1$ and disequilibrium thermochemistry. Different chemical species respond differently to both effects, and if we compare species' sensitivities to these effects, \ce{CO}, \ce{H2O} and \ce{CO2} are more sensitive to the metallicity increase, while \ce{CH4}, \ce{NH3} and \ce{HCN} are sensitive to both.

As metallicity increases, both equilibrium and kinetics simulations predict that the abundance of \ce{CO}, \ce{H2O} and \ce{CO2} increases by up to two orders of magnitude throughout the entire atmosphere. However, the spatial distribution of these species does not appreciably change: \ce{CO} and \ce{H2O} stay predominantly vertically and horizontally uniform, while \ce{CO2} follows the temperature structure by being more abundant at lower temperatures.

\begin{figure*}
 \includegraphics[scale=0.58]{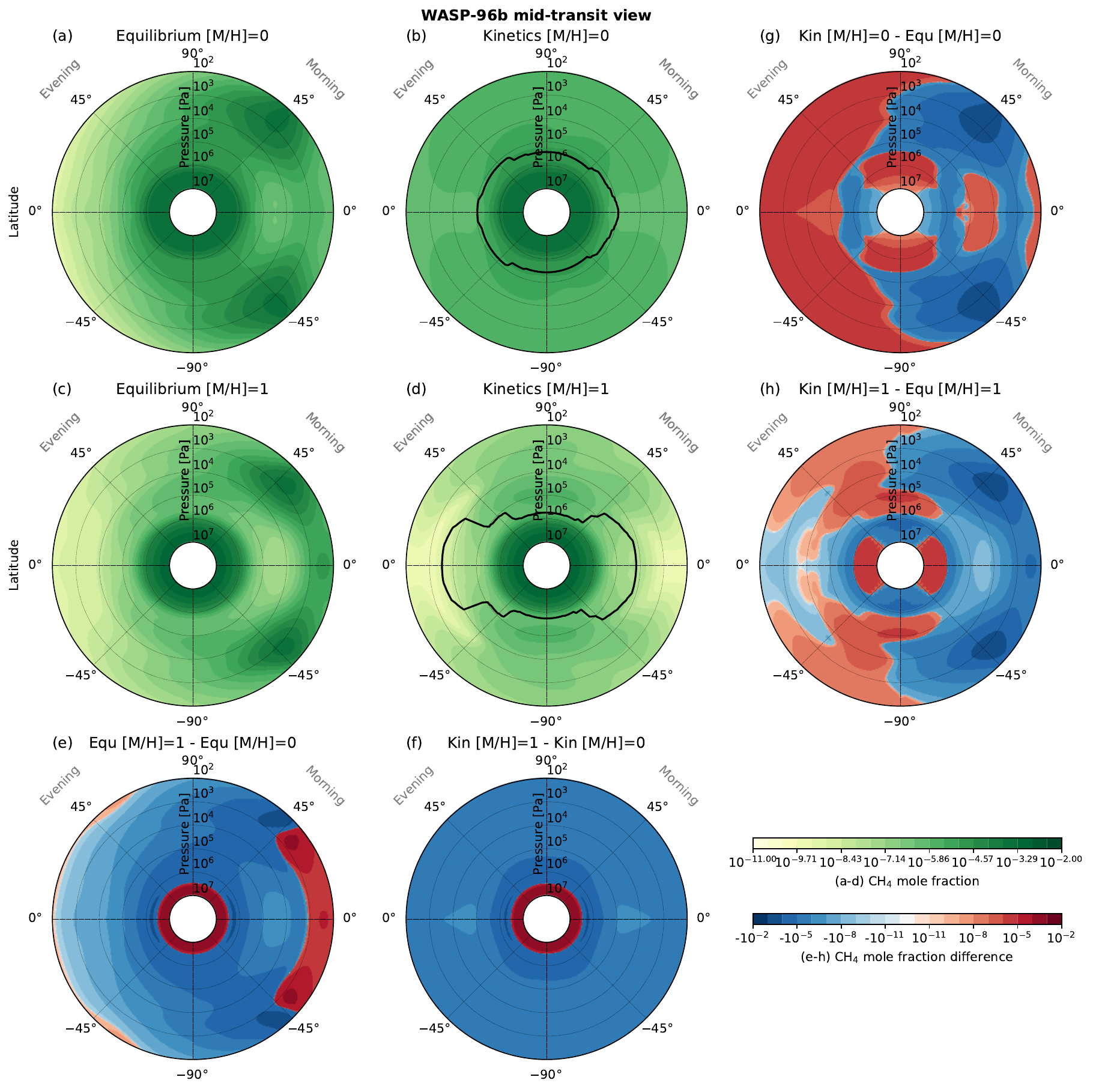}
 \caption{Pressure-\ce{CH4} cross-sections through WASP-96b's limb (``mid-transit view'') according to the equilibrium and kinetics simulations assuming $[M/H]=0$ or $[M/H]=1$. Panels (a-d) show \ce{CH4} distribution in each simulation. Panels (e-f) show the difference caused by an increase in metallicity, and (g-h) the difference caused by disequilibrium thermochemistry. Panels (a-d) and (e-h) share the top and bottom colour scale in the bottom right of the figure, respectively. The black thick lines in panels (b) and (d) indicate the location of \ce{CH4} quench level. All panels show variable distributions at longitudes \ang{90}E and \ang{270}E exactly (not the average over the opening angle).}
\label{fig:wasp96b_pres_ch4_in_transit}
\end{figure*}

The response of \ce{CH4}, \ce{NH3} and \ce{HCN} to the metallicity increase and disequilibrium thermochemistry is more complex (Fig.~\ref{fig:wasp96b_hwind_ch4_at_plev} for \ce{CH4} at \SI{e3}{\pascal}). As metallicity increases, the equilibrium simulations predict a decrease in \ce{CH4} on the dayside and the evening limb and an increase on the nightside and the morning limb (Fig.~\ref{fig:wasp96b_hwind_ch4_at_plev}e). For \ce{NH3} and \ce{HCN}, however, the equilibrium simulations predict an increase in their abundance throughout the entire atmosphere, with this increase being larger on the nightside and the morning limb, similarly to \ce{CH4}. Contrary to the equilibrium simulations, the kinetics simulations predict an overall decrease in \ce{CH4}, \ce{NH3} and \ce{HCN} at $\leq$\num{e5} \unit{\pascal} when metallicity is increased (Fig.~\ref{fig:wasp96b_hwind_ch4_at_plev}f). Meanwhile, disequilibrium thermochemistry alters \ce{CH4} distribution away from that at chemical equilibrium differently at different metallicities. At $[M/H]=0$, disequilibrium thermochemistry increases \ce{CH4} abundance outside of the Rossby gyres whist decreasing it inside the gyres, overall leading to a more horizontally homogeneous \ce{CH4} distribution (Fig.~\ref{fig:wasp96b_hwind_ch4_at_plev}b). At $[M/H]=1$, however, the same zonal homogenisation is occurring (Fig.~\ref{fig:wasp96b_hwind_ch4_at_plev}d), but the equatorial depletion in \ce{CH4} barely noticeable at $[M/H]=0$ (better seen in Fig.~\ref{fig:wasp96b_pres_ch4_in_transit}b) becomes more prominent at $[M/H]=1$. This phenomenon, the equatorial depletion of \ce{CH4}, also occurring for \ce{NH3} and \ce{HCN} (Figs.~\ref{fig:wasp96b_pres_nh3_in_transit}-\ref{fig:wasp96b_pres_hcn_in_transit}b,d), is caused by quenching.

\begin{figure*}
\includegraphics[scale=0.54]{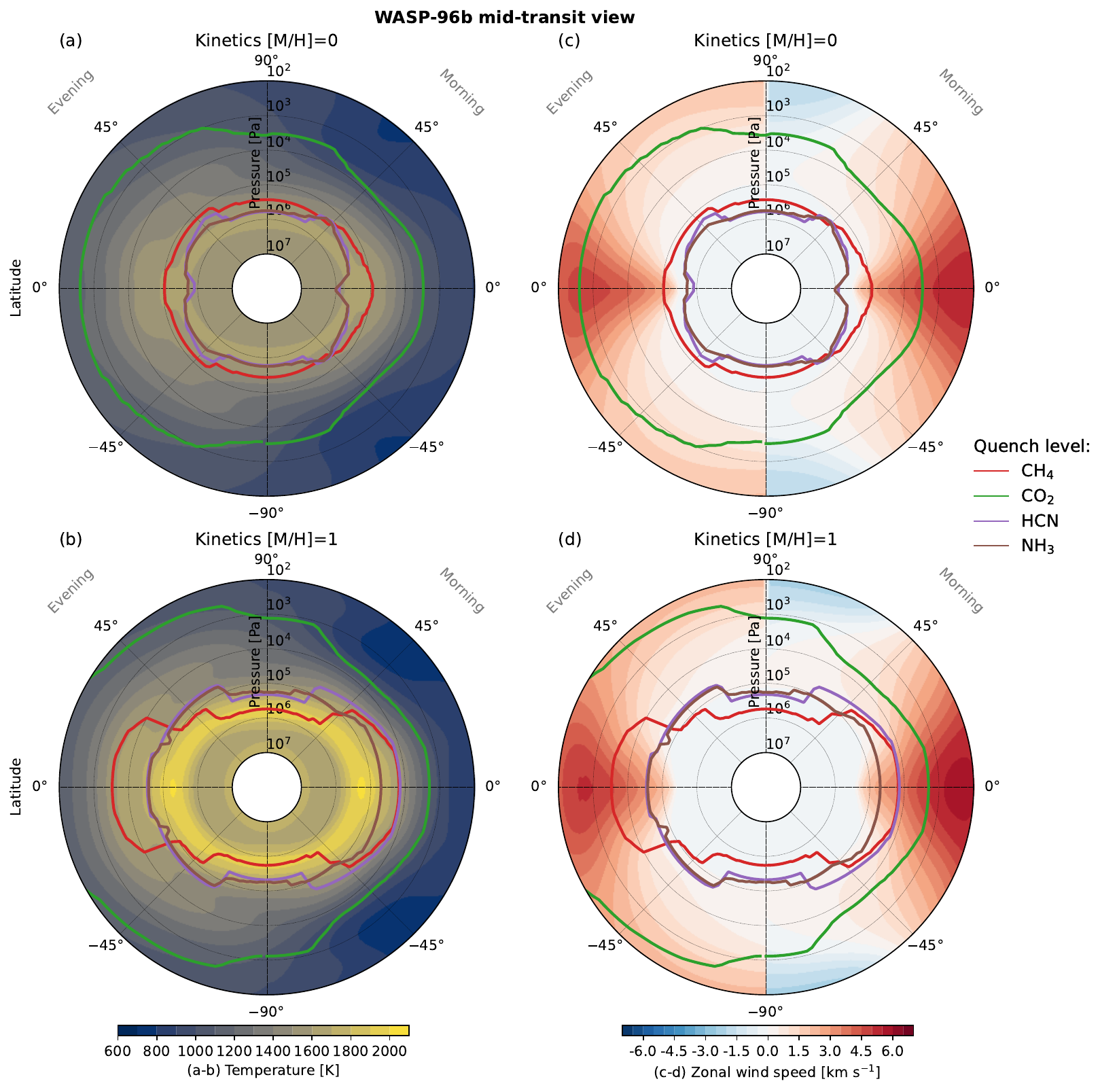}
 \caption{Pressure-temperature cross-sections through WASP-96b's terminator (``mid-transit view'') in the kinetics simulations assuming $[M/H]=0$ or $[M/H]=1$. Left column shows the thermal structure in each simulation, and right column shows zonal wind speed. Coloured lines show the location of chemical species' quench points: \ce{CH4} in red, \ce{CO2} in green, \ce{HCN} in purple and \ce{NH3} in brown. All panels show variable distributions at longitudes \ang{90}E and \ang{270}E exactly (not the average over the opening angle).}
 \label{fig:wasp96b_midtransit_temp_u_w_all_sps_qplevs}
\end{figure*} 

\subsubsection{Quenching-driven equatorial depletion}
\label{sec:equatorial_depletion}

To understand the mechanism of \ce{CH4}, \ce{NH3} and \ce{HCN} quenching-driven equatorial depletion, we manually identified the pressure level at which each of these species quenches for the first time, i.e. has their deepest quench level. To illustrate the mechanism and relate the quench level information to observations in primary transit, we show the \ce{CH4}, \ce{NH3}, \ce{HCN} (and \ce{CO2}) quench level positions in the terminator plane during mid-transit overlaid onto the temperature, zonal wind (Fig.~\ref{fig:wasp96b_midtransit_temp_u_w_all_sps_qplevs}) and chemical species abundance fields (Fig.~\ref{fig:wasp96b_pres_ch4_in_transit}b,d for \ce{CH4} and Figs.~\ref{fig:wasp96b_pres_nh3_in_transit}-\ref{fig:wasp96b_pres_co2_in_transit}b,d for other species). By presenting the data this way we effectively show where the part of the atmosphere geometrically seen in mid-transit splits into two regions from the point of view of each of these three species. One region experiences chemical equilibrium and is located inside the quench-level contours (at higher pressures), and the other region experiences thermochemical disequilibrium and is located outside the quench-level contours (at lower pressures).

\ce{CH4}, \ce{NH3} and \ce{HCN} quench at different pressures at different metallicities, and the pressure level at which their quenching occurs changes with latitude. At $[M/H]=0$, all three species quench at \SIrange{\sim e5}{e6}{\pascal}, across the entire terminator plane. At $[M/H]=1$, however, their equatorial to mid-latitude (within about $\pm$\ang{45} latitude) quench levels are located at lower pressures, \SIrange{\sim e3}{e4}{\pascal}, than elsewhere, which places these quench levels closer to the region of the equatorial jet. Such a protrusion of quench levels into the jet region suggests that the time scales of dynamical mixing ($\tau_{dyn}$) and species' chemical conversion ($\tau_{chem}$), which when in balance cause quenching \citep[e.g.,][]{Moses2011}, change with latitude. If we consider $\tau_{dyn}$ only in the zonal direction (as is shown in Fig.~\ref{fig:wasp96b_midtransit_temp_u_w_all_sps_qplevs}c,d), in the $[M/H]=1$ kinetics simulation $\tau_{dyn}$ shortens equatorward and with decreasing pressure following the increase in the zonal wind speed, while $\tau_{chem}$ reacts to changes in the speed of the prevailing species' conversion pathways.

The factor that enables and strengthens \ce{CH4}, \ce{NH3} and \ce{HCN} quenching-driven equatorial depletion is the increase in temperature at \SIrange{\sim e4}{e7}{\pascal} due to an increase in atmospheric metallicity. These pressure levels of WASP-96b's simulated atmosphere are hotter at $[M/H]=1$ than at $[M/H]=0$, which pushes isotherms critical for \ce{CH4}, \ce{NH3} and \ce{HCN} chemistry, \SIrange{\sim 1400}{1600}{\kelvin} \citep{Moses2011,Baeyens2021}, to lower pressures into the region of the equatorial jet. This, in turn, exposes gas close to chemical equilibrium to a sharp gradient in the zonal wind speed, forcing species' quench levels, i.e. the location at which $\tau_{dyn}$ equals each of the species' $\tau_{chem}$, to shift to lower pressures. The $[M/H]=1$ kinetics case additionally shows that these quench level shifts required to achieve $\tau_{dyn}=\tau_{chem}$ not only change with latitude, but also differ between the limbs. The \ce{CH4} quench levels protrude more into the jet region at the evening limb than at the morning limb, while \ce{NH3} and \ce{HCN} quench levels do the opposite. This leads to another phenomenon, \ce{CH4}, \ce{NH3} and \ce{HCN} quenching-driven limb asymmetry in equatorial depletion, which is the asymmetry in the magnitude and spatial extent of said depletion between the limbs.

The signs of \ce{CH4}, \ce{NH3} and \ce{HCN} quenching-driven equatorial depletion were present in the UM simulations of HD~189733b and HD~209458b \citep{Drummond2020,Zamyatina2023} performed with the \citet{Venot2019} chemical network, and recently were also reported to be present in the Exo-FMS simulations of HD~189733b and WASP-39b \citep{Lee2023} performed with the \citet{Tsai2022} ``mini-chemical'' network. There \textit{are} differences between the UM and Exo-FMS HD~189733b \ce{CH4}, \ce{NH3} and \ce{HCN} quenched distributions, and while these particular simulations are not directly comparable (because Exo-FMS simulations did not have the coupling between the kinetics-driven changes in species abundances and radiative transfer), in the future it is important to explore the process of quenching with different GCMs and estimate the ``GCM uncertainty error bar'' \citep{Fauchez2022} for this process.

\subsubsection{Quenching-driven limb asymmetries}
\label{sec:limb_asymmetries}

Limb asymmetry is an asymmetry in characteristics of a planetary atmosphere between parts of its western and eastern hemispheres observable in primary transit. The western part is often referred to as the morning (or leading) limb, as it is the region where atmospheric characteristics adjust from night to day conditions, while the eastern part, or the evening (or trailing) limb, is where they adjust from day to night conditions. Limb asymmetries were first detected in the atmospheres of the ultra-hot Jupiters WASP-76b and WASP-189b using high-resolution cross-correlation spectroscopy \citep{Ehrenreich2020,Prinoth2022,Kesseli2022}, and were recently found in the atmosphere of the hot Jupiter WASP-39b via the analysis of its \textit{JWST} transit light curves \citep[][Espinoza et al. (in review), Delisle et al. (submitted)]{Rustamkulov2023}. 

To correctly interpret observations of limb asymmetry, one needs to consider the drivers of limb asymmetry \citep[e.g.,][]{Savel2023}. One of the primary drivers of limb asymmetry is the temperature contrast between the limbs. The higher temperatures of the evening limb cause its atmospheric scale height to be larger than that of the morning limb, which leads to the evening-limb's transit depths being larger than those of the morning limb \citep{Pluriel2023}. Another driver of limb asymmetry is the opacity contrast between the limbs. This opacity contrast could be caused by a multitude of processes, e.g., cloud \citep{Line2016,Powell2019} or haze formation \citep{Steinrueck2021} or photochemistry \citep{Tsai2023}. In this study, we report how quenching changes the planet's limb opacity from that at chemical equilibrium, and how the associated evening-morning limb opacity contrast also changes depending on the assumed metallicity. For transmission spectra, we specifically focus on the mid-transit. We use the methodology of \citet{Lines2018} and \citet{Christie2021} to calculate the full, morning and evening limb transmission spectra, with the morning and evening limb spectra computed in the same way as the full limb spectra except that the given limb is mirrored to provide an appropriate planetary radius for the whole planet. The abundances of radiatively active species, \ce{H2O}, \ce{CO}, \ce{CO2}, \ce{CH4}, \ce{NH3}, \ce{HCN}, \ce{Li}, \ce{Na}, \ce{K}, \ce{Rb}, \ce{Cs}, \ce{H2} and \ce{He}, are taken directly from the simulations, allowing for the abundances used in transmission spectra post-processing to be consistent with the chemistry used in the simulations.

According to our simulations, quenching alters the limb distribution of all chemical species, including those of \ce{CH4}, \ce{CO}, \ce{CO2}, \ce{H2O}, \ce{NH3} and \ce{HCN} (Figs.~\ref{fig:wasp96b_pres_ch4_in_transit}g-h, \ref{fig:wasp96b_pres_nh3_in_transit}-\ref{fig:wasp96b_pres_co_in_transit}g-h, with vertical profiles shown in Figs.~\ref{fig:wasp96b_vp_pres_ch4_limbs}-\ref{fig:wasp96b_vp_pres_nh3_limbs}). While the magnitude and location of these alterations varies amongst species, the general trend is that species that quench deeper in the atmosphere (e.g., \ce{CH4}, \ce{NH3}, \ce{HCN}), as opposed to those that quench in the upper atmosphere (e.g., \ce{CO2}) or do not quench at all (e.g., \ce{OH} or any other free radical), have a substantially more zonally-uniform limb distribution than that predicted at chemical equilibrium. For example, in the equilibrium simulations \ce{CH4} is abundant in the cold Rossby gyres at the morning limb and is depleted in the warmer upper-atmosphere regions elsewhere, while in the kinetics simulations \ce{CH4} limb distribution is zonally uniform, including the region of its equatorial depletion. In contrast to \ce{CH4}, \ce{CO2} quenched limb distribution is similar to that at chemical equilibrium. This is because \ce{CO2} quenches at much lower pressures than \ce{CH4}, which leaves a large part of \ce{CO2} limb distribution untouched by quenching, especially at the morning limb, and instead determined by chemical equilibrium.

\begin{figure*}
 \includegraphics[scale=0.62]{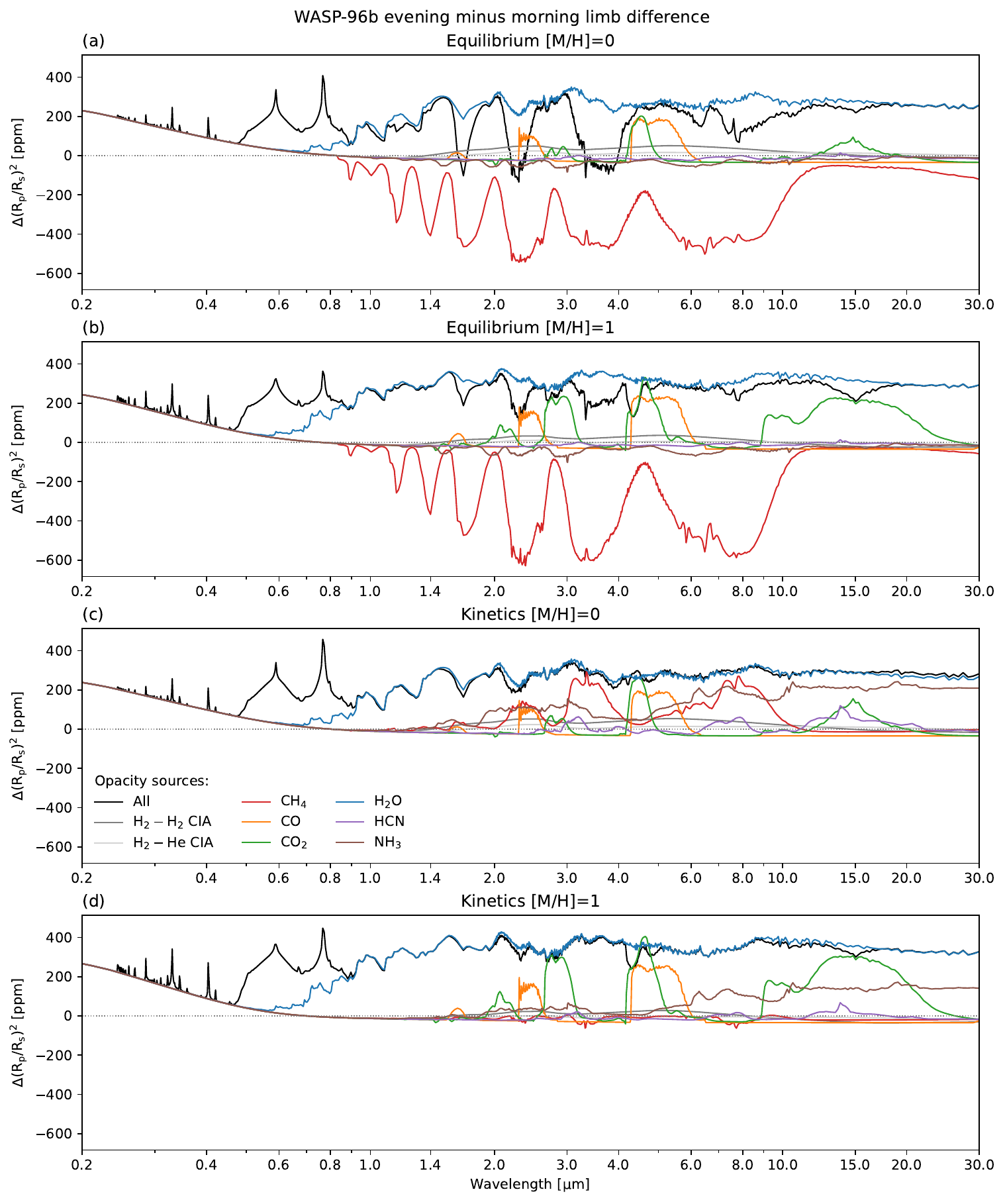}
 \caption{Contributions of individual opacity sources to the difference between evening and morning limb transmission spectra (evening minus morning) according to the WASP-96b equilibrium and kinetics simulations assuming $[M/H]=0$ or $[M/H]=1$. Faint grey dotted lines indicate the location of zero to better guide the eye.}
 \label{fig:wasp96b_transpec_wl02_30_combores_eve_m_mor}
\end{figure*}

To understand if quenching-driven changes in the limb distributions of \ce{CH4}, \ce{NH3}, \ce{HCN} and \ce{CO2} could help distinguish aerosol-free atmospheres at chemical equilibrium from those at thermochemical disequilibrium, we calculated the evening minus morning limb transit depth differences caused by all and individual opacity sources separately (Fig.~\ref{fig:wasp96b_transpec_wl02_30_combores_eve_m_mor}, with limb transit depth differences due to all opacity sources shown together in Fig.~\ref{fig:wasp96b_transpec_wl02_30_combores_full_mor_eve}d). We find that these evening-morning limb differences vary with wavelength and across simulations. The main contributors to these differences are Rayleigh scattering (shortward of $\sim$\SI{0.5}{\micro\metre}), \ce{Na} and \ce{K} absorption (within $\sim$\SIrange{0.5}{0.9}{\micro\metre}), and \ce{H2O} absorption (longward of $\sim$\SI{0.9}{\micro\metre}). All the main contributors have a higher opacity at the evening limb than at the morning limb due to a higher temperature at the evening limb. Meanwhile, the contribution of other opacity sources to the evening-morning limb differences changes between simulations. The contribution of \ce{CH4} changes the most. In the equilibrium simulations, \ce{CH4} has an opposite in sign and comparable to \ce{H2O} limb contrast, which is large enough to diminish the cumulative limb contrast of all opacity sources at wavelengths where \ce{CH4} strongly absorbs. In the kinetics simulations, however, \ce{CH4} limb contrast is mostly smaller than that of \ce{H2O}, yet it changes with metallicity: at $[M/H]=0$ \ce{CH4} is abundant and there is more of it at the evening limb, while at $[M/H]=1$ \ce{CH4} is depleted and there is a similar amount of it at both limbs. At around $\sim$\SI{3.32}{\micro\metre}, \ce{CH4} strongest absorption line in the infrared, the \ce{CH4} limb contrast in the $[M/H]=0$ kinetics simulation is comparable to that of \ce{H2O}, which causes the cumulative limb contrast of all opacity sources to increase around that wavelength.

The limb distribution of \ce{CO2} also changes enough amongst our simulations to modify the evening-morning limb transit depth differences. Specifically, if we compare these differences at around $\sim$\SI{4.23}{\micro\metre}, \ce{CO2} strongest absorption line in the infrared, $[M/H]=0$ simulations suggest that the abundance of \ce{CO2} is low enough comparative to that of \ce{H2O}, so that the \ce{CO2} limb contrast does not modify the cumulative limb contrast. However, $[M/H]=1$ simulations show that the \ce{CO2} abundance becomes high enough to be comparable to that of \ce{H2O}, and as a result the \ce{CO2} limb contrast causes a dip in the cumulative limb contrast at around $\sim$\SI{4.23}{\micro\metre}.

\subsection{Parts of transmission spectrum sensitive to metallicity and quenching}
\label{sec:sensitivity_to_mdh_and_quenching}

\begin{figure*}
 \includegraphics[scale=0.62]{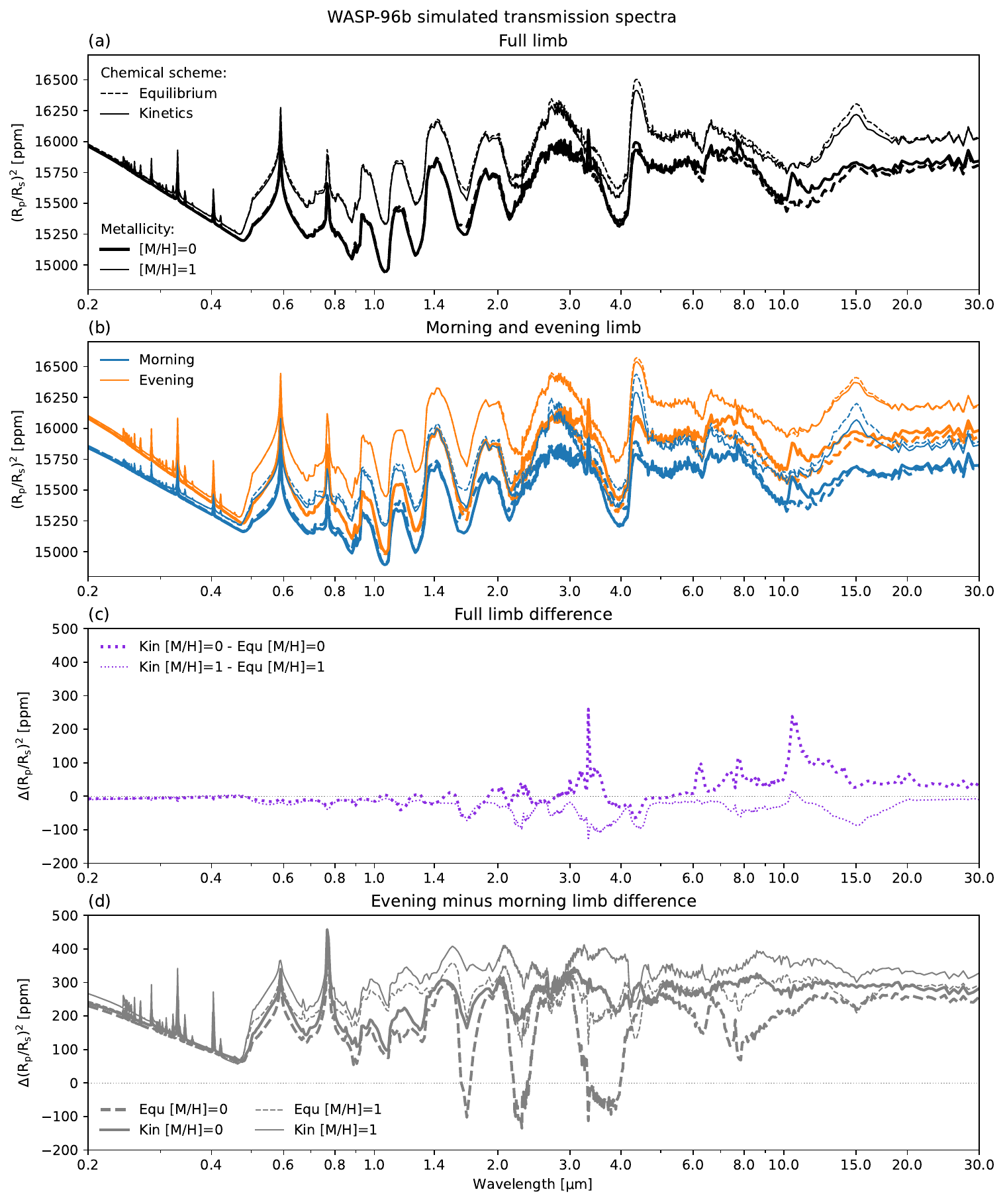}
 \caption{WASP-96b transmission spectra predicted by the Met Office \textsc{Unified Model} equilibrium and kinetics simulations assuming $[M/H]=0$ or $[M/H]=1$. Panel (a) shows the full limb (morning plus evening) spectra, and (b) the morning and evening limb spectra. Panel (c) shows the difference between the full limb spectra due to the disequilibrium thermochemistry, and panel (d) shows the difference between evening and morning limb spectra (evening minus morning). Faint grey dotted lines indicate the location of zero to better guide the eye.}
 \label{fig:wasp96b_transpec_wl02_30_combores_full_mor_eve}
\end{figure*}

Metallicity- and quenching-driven changes in the predicted atmospheric structure of WASP-96b translate into a number of differences in the planet's simulated transmission spectra (Figs.~\ref{fig:wasp96b_transpec_wl02_30_combores_eve_m_mor}-\ref{fig:wasp96b_transpec_wl02_30_combores_full_mor_eve}, \ref{fig:wasp96b_transpec_wl02_30_combores_contribs_to_full}-\ref{fig:wasp96b_transpec_wl02_30_combores_mor_eve}). To discuss these differences, we show our simulated spectra in five more different ways: (1) the full limb (morning plus evening) spectra (Fig.~\ref{fig:wasp96b_transpec_wl02_30_combores_full_mor_eve}a), (2) the morning and evening limb spectra (Figs.~\ref{fig:wasp96b_transpec_wl02_30_combores_full_mor_eve}b, \ref{fig:wasp96b_transpec_wl02_30_combores_mor_eve}), (3) the difference in the full limb spectra due to disequilibrium thermochemistry (Fig.~\ref{fig:wasp96b_transpec_wl02_30_combores_full_mor_eve}c), (4) the difference between evening and morning limb spectra (Fig.~\ref{fig:wasp96b_transpec_wl02_30_combores_full_mor_eve}d), and (5) the contributions of individual opacity sources to the full limb spectra (Fig.~\ref{fig:wasp96b_transpec_wl02_30_combores_contribs_to_full}).


In the case of the full limb spectra, spectral differences caused by disequilibrium thermochemistry occur longward of the \ce{H2}- and \ce{He}-induced Rayleigh scattering slope, with the largest difference being at around $\sim$\SI{3.32}{\micro\metre}. Around that particular wavelength, the quenching-driven changes in \ce{CH4}, an enhancement at $[M/H]=0$ and a depletion at $[M/H]=1$, causes transit depths from $[M/H]=0$ kinetics simulation to be $\sim$300 ppm \textit{larger} than those from $[M/H]=0$ equilibrium simulation (Fig.~\ref{fig:wasp96b_transpec_wl02_30_combores_full_mor_eve}c). Meanwhile, the corresponding transit depths from $[M/H]=1$ kinetics simulation are $\sim$100 ppm \textit{smaller} than those from $[M/H]=1$ equilibrium simulation. Such a difference in the sign of the change in \ce{CH4} \SI{3.32}{\micro\metre} feature with increasing metallicity due to disequilibrium thermochemistry has two consequences. One, the \ce{CH4} \SI{3.32}{\micro\metre} feature is a more powerful gauge of metallicity for hot Jupiter atmospheres at thermochemical disequilibrium than for atmospheres at chemical equilibrium. Two, for WASP-96b the presence of that feature in the planet's full limb transmission spectrum (Fig.~\ref{fig:wasp96b_transpec_wl02_30_combores_full_mor_eve}a) would support a scenario of a lower, $[M/H]=0$ metallicity atmosphere at thermochemical disequilibrium, rather than one of the other three scenarios considered in this study. The same argument is true for \ce{CH4} \SI{7.70}{\micro\metre} and \ce{NH3} \SI{10.37}{\micro\metre} features, but the metallicity- and quenching-driven changes in these features are smaller relative to those of \ce{CH4} at \SI{3.32}{\micro\metre}. \ce{CO2} \SI{4.23}{\micro\metre} and \SI{14.95}{\micro\metre} features, while also affected by the metallicity change and quenching, are less powerful in uniquely characterising hot Jupiter atmospheric metallicity. \ce{CO2} \SI{4.23}{\micro\metre} feature is present in all our scenarios and only the size of the feature is modified (Fig.~\ref{fig:wasp96b_transpec_wl02_30_combores_contribs_to_full}). \ce{CO2} \SI{14.95}{\micro\metre} feature, while present only at $[M/H]=1$, is broad and is competing with a strong \ce{H2O} absorption. That means that putting information about \ce{CO2} features into the context of surrounding wavelengths is critical for a correct identification of atmospheric metallicity.

In the case of the evening-morning limb asymmetries, spectral differences caused by the metallicity increase from $[M/H]=0$ to $[M/H]=1$ and disequilibrium chemistry are comparable in size. The evening limb transit depths are $\sim$100-400 ppm larger than those of the morning limb in all our simulations except for the $[M/H]=0$ equilibrium one (Fig.~\ref{fig:wasp96b_transpec_wl02_30_combores_full_mor_eve}d). In that particular simulation, the transit depths for the evening limb are $\sim$100 ppm smaller than those of the morning limb around \ce{CH4} 1.67, 2.37, 3.21, 3.32 and \SI{3.42}{\micro\metre} features. This sign reversal for the evening-morning limb difference is unique to this simulation from our set of four simulations, and is caused by the fact that a higher than in any other scenario morning limb's \ce{CH4} abundance and the resulting \ce{CH4} limb contrast (Fig.~\ref{fig:wasp96b_transpec_wl02_30_combores_eve_m_mor}) outweighs the temperature limb contrast present in this simulation. A similar argument is true for \ce{CH4} 7.40 and \SI{7.70}{\micro\metre} features, but the lower pressure regions of the atmosphere sampled at these wavelengths do not have a strong enough \ce{CH4} limb contrast to counteract the temperature limb contrast. While the analysis above helps separate the $[M/H]=0$ equilibrium scenario from the other three considered in this study, distinguishing between both kinetics simulations and $[M/H]=1$ equilibrium simulation is more difficult. At most wavelengths an increase in metallicity causes an offset between the evening-morning limb differences from the respective equilibrium and kinetics simulations, while offsets are known to be challenging to interpret if wavelength coverage is non-continuous (e.g., Carter et al., in review). Another difficulty is that different scenarios could produce similar in amplitude and shape evening-morning limb differences, for example, $[M/H]=0$ kinetics and $[M/H]=1$ equilibrium simulations within $\sim$\SIrange{1.6}{3.0}{\micro\metre} and $\sim$\SIrange{5.0}{7.0}{\micro\metre} (Fig.~\ref{fig:wasp96b_transpec_wl02_30_combores_full_mor_eve}d). However, $\sim$\SIrange{3.0}{5.0}{\micro\metre} is the region where all four of our simulations predict \textit{different in amplitude and shape} evening-morning limb differences, caused by the differences in \ce{H2O}, \ce{CH4} and \ce{CO2} (see Section~\ref{sec:limb_asymmetries}). This makes the $\sim$\SIrange{3.0}{5.0}{\micro\metre} region the most powerful region for distinguishing atmospheres at chemical equilibrium from those at thermochemical disequilibrium and uniquely identifying their metallicity.

\subsection{Comparison to existing observations}
\label{sec:comparison_to_existing_obs}

\begin{figure*}
 \includegraphics[scale=0.58]{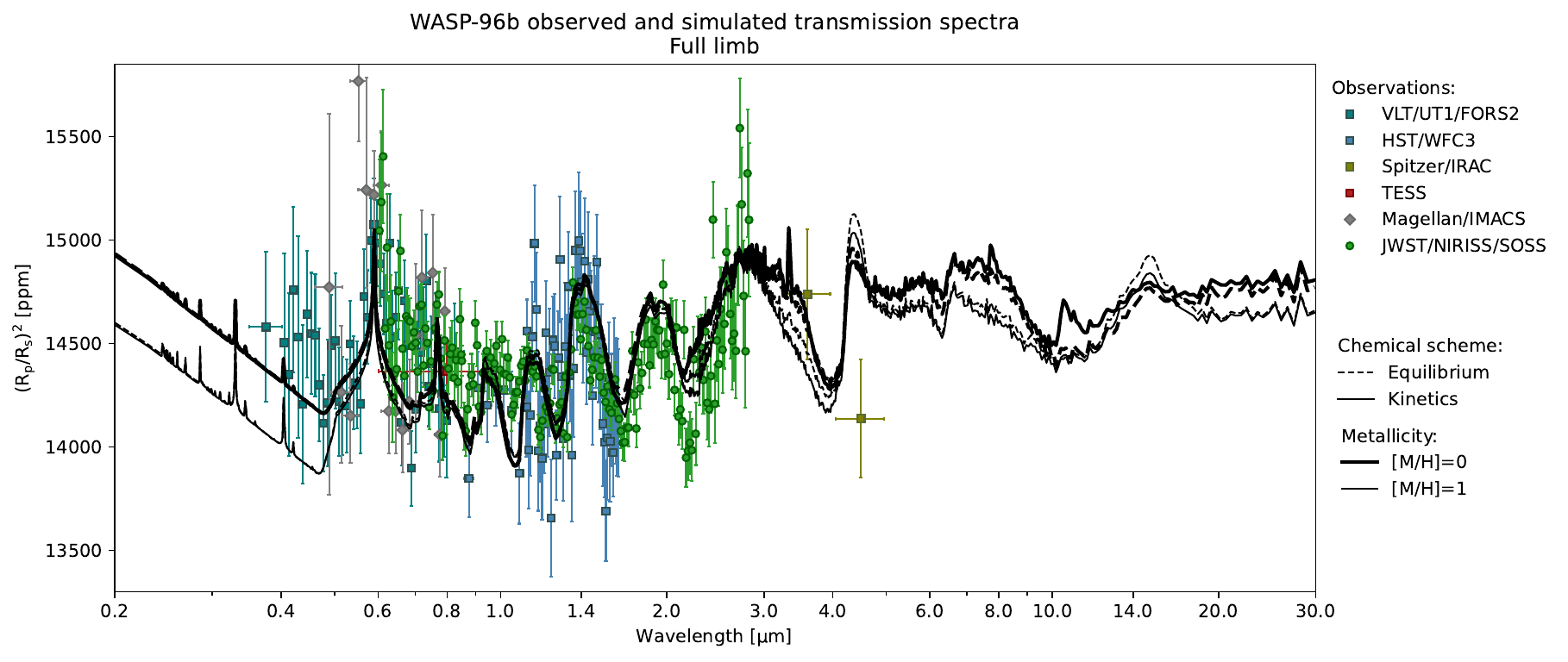}
 \caption{WASP-96b observed and simulated full limb (morning plus evening) transmission spectra. Transit depths observed with \textit{VLT}/UT1/FORS2, \textit{HST}/WFC3 G102 and G141, \textit{Spitzer}/IRAC and \textit{TESS} \citep{Nikolov2022}, \textit{Magellan}/IMACS \citep{McGruder2022} and \textit{JWST}/NIRISS/SOSS \citep{Radica2023} (the latter is shown at R$\approx$125) are overlaid with the spectra predicted by the Met Office \textsc{Unified Model} equilibrium and kinetics simulations assuming $[M/H]=0$ or $[M/H]=1$.}
 \label{fig:wasp96b_transpec_wl02_30_combores_fit_to_obs}
\end{figure*}

We compare an ensemble of the observed transmission spectra of WASP-96b from \citet{Nikolov2022}, \citet{McGruder2022} and \citet{Radica2023} (the latter at R$\approx$125) to our simulated spectra (Fig~\ref{fig:wasp96b_transpec_wl02_30_combores_fit_to_obs}). Following \citet{Radica2023}, we offset \citet{Nikolov2022}'s \textit{VLT}/UT1/FORS2 and \textit{HST}/WFC3 G102 and G141 spectra relative to the \textit{JWST}/NIRISS/SOSS spectrum by +400 ppm and +200 ppm, respectively. We also offset \citet{McGruder2022}'s \textit{Magellan}/IMACS-only spectrum by +1200 ppm relative to \textit{JWST}/NIRISS/SOSS. Using the least squares method and the combined \citet{Nikolov2022}'s \textit{VLT}/UT1/FORS2 and \textit{HST}/WFC3 G102 and G141 and \citet{Radica2023}'s \textit{JWST}/NIRISS/SOSS data, we fitted for offsets between the observed and simulated spectra, with these offsets resulting to be -1035 ppm and -1378 ppm for $[M/H]=0$ and $[M/H]=1$ pairs of simulations, respectively.

None of our simulated spectra match the existing observations perfectly, however this imperfection reveals the following:
\begin{itemize}
    \item[(1)] The \SIrange{1.30}{2.15}{\micro\metre} region is the only region where the transit depths predicted by all \textsc{UM} simulations stay mostly within the observational uncertainties. This is also the region where all simulated spectra differ the least, and show that the main contributor to the opacity is \ce{H2O}. Both of these factors suggest that all our simulations are consistent with WASP-96b's atmosphere having \ce{H2O}.
    \item[(2)] For \SIrange{0.35}{0.60}{\micro\metre}, the spectra from our $[M/H]=0$ simulations agree better with the \textit{VLT}/UT1/FORS2 observations, however because we applied an offset of +400 ppm to the \textit{VLT}/UT1/FORS2 data further interpretation is difficult.
    \item[(3)] The \SIrange{0.60}{1.30}{\micro\metre} region is the region where \ce{K} and \ce{H2O} absorption features predicted by the UM are not muted enough compared to observations. This is consistent with the presence of a scattering haze in WASP-96b's atmosphere.
    \item[(4)] The \SIrange{2.15}{2.80}{\micro\metre} region is the region where the observed and all our simulated spectra broadly agree. However, because the observed transit depths vary a lot with wavelength, the identification of the best theoretical scenario is difficult.
    \item[(5)] At \SI{3.6}{\micro\metre}, all UM simulations predict transit depths that are within the uncertainty of the transit depth observed in this \textit{Spitzer}/IRAC channel.
    \item[(6)] At \SI{4.5}{\micro\metre}, all UM simulations predict larger transit depths than the one observed in this \textit{Spitzer}/IRAC channel.
\end{itemize}

Overall, our simulated WASP-96b transmission spectra are rather similar at the wavelengths that have been already observed, so observing WASP-96b at longer wavelengths, where our predicted spectra diverge more, would help better constrain WASP-96b's atmospheric metallicity.

\section{Discussion}
\label{sec:discussion}

The question of observability of the evening-morning limb asymmetries in hot Jupiter atmospheres has been of interest for some time from both the theoretical \citep[e.g.,][]{Fortney2010,Kempton2017,Powell2019,Helling2020,MacDonald2020} and observational \citep[e.g.,][]{VonParis2016,Espinoza2021,Grant2023} points of view. Early GCM aerosol-free simulations of \citet{Fortney2010} of the atmosphere of the hot Jupiter HD~189733b predicted that at $[M/H]=0$ and chemical equilibrium the morning and evening hemisphere spectra of that planet would be different, with \ce{CH4} absorption features being larger in the morning hemisphere spectra than the evening hemisphere spectra. When the GCM used in \citet{Fortney2010} included a simple representation of disequilibrium thermochemistry through the chemical relaxation scheme of \citet{Cooper2006}, however, the spectra of both hemispheres became almost identical, with large \ce{CH4} absorption features being present in both hemispheres' spectra due to the homogenisation of \ce{CH4} abundances by vertical mixing. Their results qualitatively agree with our results from WASP-96b $[M/H]=0$ and $[M/H]=1$ equilibrium and $[M/H]=0$ kinetics simulations, respectively. However, our $[M/H]=1$ kinetics simulation additionally suggests that while the spectra of both limbs are almost identical due to the homogenisation of quenched species abundances by horizontal and vertical mixing, quenching-driven equatorial depletion of \ce{CH4} causes \ce{CH4} absorption features in both limbs' spectra to be small and possibly undetectable (Fig.~\ref{fig:wasp96b_transpec_wl02_30_combores_mor_eve}). The impact of aerosols on the limbs' spectra of hot and ultra-hot Jupiters was explored by \citet{Kempton2017} and \citet{Powell2019}, who also estimated the size of the combined effect of the temperature and aerosol opacity contrast between the limbs on the limbs' spectra. \citet{Kempton2017}'s simulations with a GCM coupled (but not radiatively) to a phase-equilibrium cloud model predicted that at $[M/H]=0$ the differences between the morning and evening limb spectra of the ultra-hot Jupiter WASP-121b could be $\sim$100 ppm at $\sim$\SIrange{0.7}{5.0}{\micro\metre}, rising up to 400 ppm at optical wavelengths in the case of a purely Rayleigh scattering aerosol. \citet{Powell2019}'s simulations with a size-distribution-resolving cloud microphysics 1D model suggested that for hot Jupiters with equilibrium temperatures of \SIrange{1800}{2100}{\kelvin} and $[M/H]=0$, limb differences could be $\sim$600 ppm at $\sim$\SIrange{0.7}{30.0}{\micro\metre}, rising up to $\sim$\SIrange{700}{1200}{ppm} in the optical. Our aerosol-free WASP-96b GCM simulations forecast similar in magnitude evening-morning limb differences, $\sim$\SIrange{100}{400}{ppm} at \SIrange{0.2}{30}{\micro\metre}, to those of \citet{Kempton2017} and \citet{Powell2019}, with quenching contributing up to $\sim$300 ppm at $[M/H]=0$ and $\sim$200 ppm at $[M/H]=1$ at $\sim$\SIrange{2.0}{30.0}{\micro\metre} to these differences (Fig.~\ref{fig:wasp96b_transpec_wl02_30_combores_full_mor_eve}d). From an observational perspective, several studies proposed techniques for detecting limb inhomogeneities. \citet{VonParis2016} suggested inferring these inhomogeneities from transit light curves by modelling the signatures of morning and evening limbs imprinted onto these light curves as two stacked semicircles with different radii. This technique was later expanded upon by \citet{Espinoza2021}, who also performed a detailed study into the observational prospects of detecting limb inhomogeneities with that technique when applied to observations with \textit{TESS} and \textit{JWST}. \citet{Espinoza2021} reported that the asymmetry in the \textit{JWST}/NIRISS/SOSS transit light curves is expected to be detectable for the evening-morning transit depth differences above $\sim$25 ppm for a wide range of spectrophotometric precisions. Finally, \citet{Grant2023} recently proposed another technique capable of inferring limb inhomogeneities from transit light curves, called transmission strings. This technique enables the extraction of transmission spectra not only for the morning and evening limbs separately, but also as a function of angle around a planet's terminator, i.e., planet's latitude. Latitudinally-resolved transmission spectra could be especially helpful for finding signs of species' quenching-driven equatorial depletion as well as of inhomogeneities in limbs' aerosol coverage. Overall, given the predictions from theory and the capabilities of existing observational techniques, there is potential for detecting evening-morning limb asymmetries in the atmosphere of WASP-96b and other similar planets.

If we compare \ce{H2O}, \ce{CO} and \ce{CO2} abundances retrieved from the \textit{JWST}/NIRISS/SOSS transmission spectrum and assumed to be representative for the full limb of WASP-96b at $\sim$\SIrange{e3}{e1}{\pascal} \citep{Taylor2023} with \ce{H2O}, \ce{CO} and \ce{CO2} abundances predicted for the morning and evening limbs at \SIrange{e3}{e2}{\pascal} from our WASP-96b GCM simulations (Figs.~\ref{fig:wasp96b_vp_pres_co_limbs}-\ref{fig:wasp96b_vp_pres_h2o_limbs}), we find that the retrieved \ce{H2O} abundance agrees better with \ce{H2O} abundances from our $[M/H]=0$ simulations, while the retrieved \ce{CO2} abundance agrees better with \ce{CO2} abundances from our $[M/H]=1$ simulations. The retrieved \ce{CO} abundance was poorly constrained, resulting in its mean value agreeing better with \ce{CO} abundances from our $[M/H]=0$ simulations but not ruling out higher values consistent with our $[M/H]=1$ simulations. This comparison suggests that given the limitations of their retrievals and our GCM, knowing \ce{H2O}, \ce{CO} and \ce{CO2} abundances might not be enough to constrain WASP-96b's atmospheric metallicity.

Putting our WASP-96b $[M/H]=0$ kinetics simulation into context of previous similar studies \citep{Drummond2020,Zamyatina2023}, which, importantly, considered only $[M/H]=0$, we find that our WASP-96b results are most similar to the results from their HD~189733b and HD~209458b $[M/H]=0$ kinetics simulations. This is unsurprising as the WASP-96b system parameters are more similar to the system parameters of HD~189733b and HD~209458b than to those of HAT-P-11b and WASP-17b also present in \citet{Zamyatina2023}'s planet sample. However, the differences between the system parameters of WASP-96b, HD~189733b and HD~209458b alter the predictions of \ce{CH4}, \ce{NH3} and \ce{HCN} quenching behaviour for these planets in such a way that puts WASP-96b's results somewhere in between those for HD~189733b and HD~209458b. For \ce{CH4} specifically (Fig.~\ref{fig:wasp96b_vp_pres_ch4_limbs}), as the most spatially variable quenched species amongst \ce{CH4}, \ce{NH3} and \ce{HCN} in the case of these three planets, WASP-96b's meridional \ce{CH4} quenching seems to be weaker than that of HD~189733b, causing WASP-96b's \ce{CH4} abundances at $\le$\SI{e4}{Pa} to be lower than those of HD~189733b but larger than those of HD~209458b. An investigation of the drivers of such differences is reserved for a future study. 

Quenching is a well-known process in planetary sciences, especially after it was used to explain the discovery of \ce{CO} in the upper atmosphere of Jupiter \citep{Prinn1977}. Since then many theoretical studies considered quenching by either performing chemical kinetics calculations or relying on the quenching approximation \citep[e.g.,][]{Smith1998,Line2010,Madhusudhan2011,Visscher2011,Zahnle2014,Fortney2020,Soni2023,Soni2023a}. While such an approximation is simple and computationally efficient, the results obtained using this approximation often diverge from the chemical kinetics results for reasons discussed in detail in \citet{Tsai2017}. Knowing that, we hope that our first demonstration of the variability in the quench level positions derived using a GCM coupled to a chemical kinetics scheme would encourage more studies similar to ours and benefit other modelling frameworks interested in quenching.

Aerosol formation in hot Jupiter atmospheres, and in particular formation of oxygen-bearing clouds, was estimated to be able to sequester up to 30\% of oxygen from the entire atmosphere's gas phase under assumption of chemical equilibrium \citep{Lee2016a,Lee2023a}. Such a depletion in oxygen would change the abundance and distribution of, e.g., \ce{H2O} and of elemental ratios such as \ce{C}/\ce{O}. The \citet{Helling2023} classification of gaseous exoplanets suggests that WASP-96b's \ce{C}/{O} would be globally increased in cloud-forming layers due to globally homogeneous cloud coverage. Given that their modelling framework relied on chemical equilibrium calculations for the gas-phase and our gas-phase only simulations did not include clouds, currently it is hard to predict how \ce{C}/\ce{O} would change if the gas-phase and cloud-phase chemical kinetics were coupled. Answering this question is one of the most challenging ongoing pursuits in exoplanet science.

\section{Conclusions}
\label{sec:conclusions}

We investigated the sensitivity of transport-induced quenching to an increase in atmospheric metallicity from $[M/H]=0$ to $[M/H]=1$ using 3D GCM simulations of an assumed to be aerosol-free hot Jupiter WASP-96b as an example. We found that:

\begin{itemize}
\item Quenching affects the spatial distribution of all chemical species, but species that quench zonally, e.g., \ce{CH4}, \ce{NH3} and \ce{HCN}, are affected by this process the most.
\item The increase in atmospheric temperature at pressures $\sim$\SIrange{e4}{e7}{\pascal} associated with an increase in atmospheric bulk metallicity could shift the location of the quench level to lower pressures, into the region of equatorial jet, and cause an equatorial depletion of species undergoing zonal quenching.
\item The $\sim$\SIrange{3.0}{5.0}{\micro\metre} region is the most powerful region in terms of evening-morning limb asymmetries for distinguishing atmospheres at chemical equilibrium from those with upper layers at thermochemical disequilibrium, and uniquely identifying their metallicity. This occurs because the amplitude and shape of evening-morning limb differences in that region are drastically different between our $[M/H]=0$ and $[M/H]=1$ equilibrium and kinetics simulations due to the difference in the abundance of \ce{H2O} and quenching behaviour of \ce{CO2} and \ce{CH4}.
\end{itemize}

\section*{Acknowledgements}

The authors thank the anonymous referee for carefully reading our manuscript and providing insightful recommendations to improve it. We also thank Arjun Baliga Savel and Maria E. Steinrueck for helping us find an error in our post-processing code.\\
This research was supported by a UKRI Future Leaders Fellowship MR/T040866/1, and partly supported by the Leverhulme Trust through a research project grant RPG-2020-82 alongside a Science and Technology Facilities Council Consolidated Grant ST/R000395/1.\\
This research was performed using the DiRAC Data Intensive service at Leicester, operated by the University of Leicester IT Services, which forms part of the STFC DiRAC HPC Facility (\url{www.dirac.ac.uk}). The equipment was funded by BEIS capital funding via STFC capital grants ST/K000373/1 and ST/R002363/1 and STFC DiRAC Operations grant ST/R001014/1. DiRAC is part of the National e-Infrastructure.\\
Material produced using Met Office Software. The scripts to process and visualise the Met Office \textsc{Unified Model} data are available on GitHub at \url{https://github.com/mzamyatina/quenching_driven_depletion_and_asymmetries}; these scripts are dependent on the following Python libraries: \texttt{aeolus} \citep{aeolus}, \texttt{iris} \citep{iris}, \texttt{ipython} \citep{ipython}, \texttt{jupyter} \citep{jupyternotebook}, \texttt{matplotlib} \citep{matplotlib} and \texttt{numpy} \citep{numpy}.\\
We would also like to note the energy intensive nature of supercomputing, especially simulations with interactive chemistry. We estimate that the final production runs needed for this paper resulted in roughly 2.3 t\ce{CO2}e emitted into the atmosphere.

\section*{Data Availability}

The Met Office \textsc{Unified Model} data supporting this publication will be made available from the University of Exeter’s institutional repository when the paper has been accepted.\\
For the purpose of open access, the authors have applied a Creative Commons Attribution (CC BY) licence to any Author Accepted Manuscript version arising.


\bibliographystyle{mnras}
\bibliography{references}


\newpage
\appendix

\section{Appendix}

\begin{figure*}
 \includegraphics[scale=0.58]{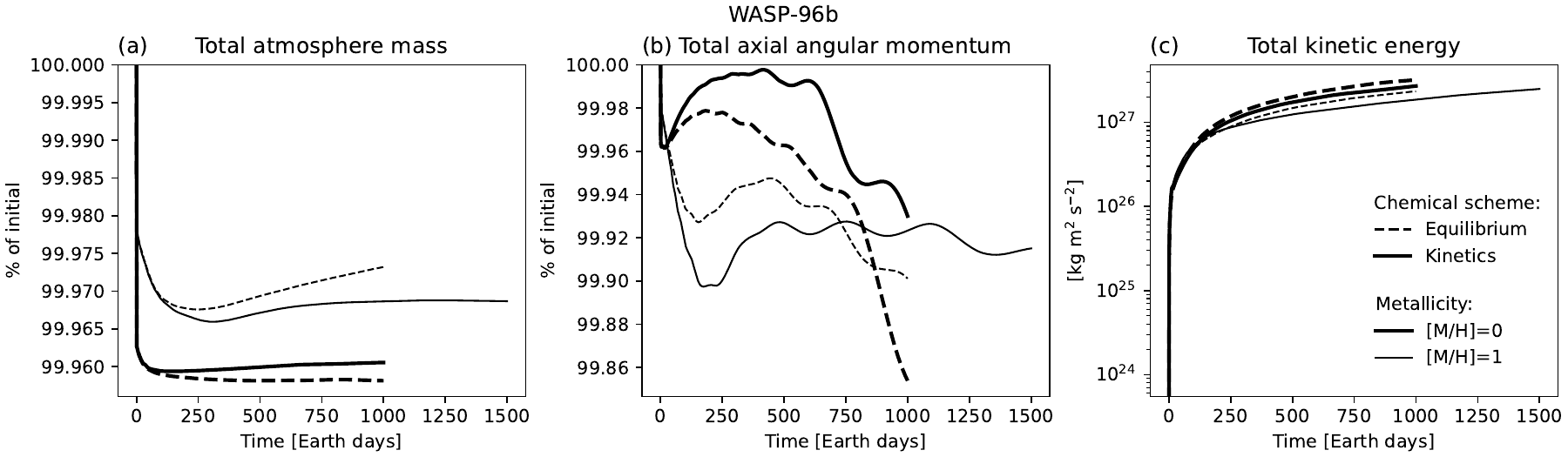}
 \caption{Conservation diagnostics from the WASP-96b equilibrium and kinetics simulations assuming $[M/H]=0$ or $[M/H]=1$.}
\label{fig:wasp96b_conservation}
\end{figure*}

\begin{figure*}
 \includegraphics[scale=0.58]{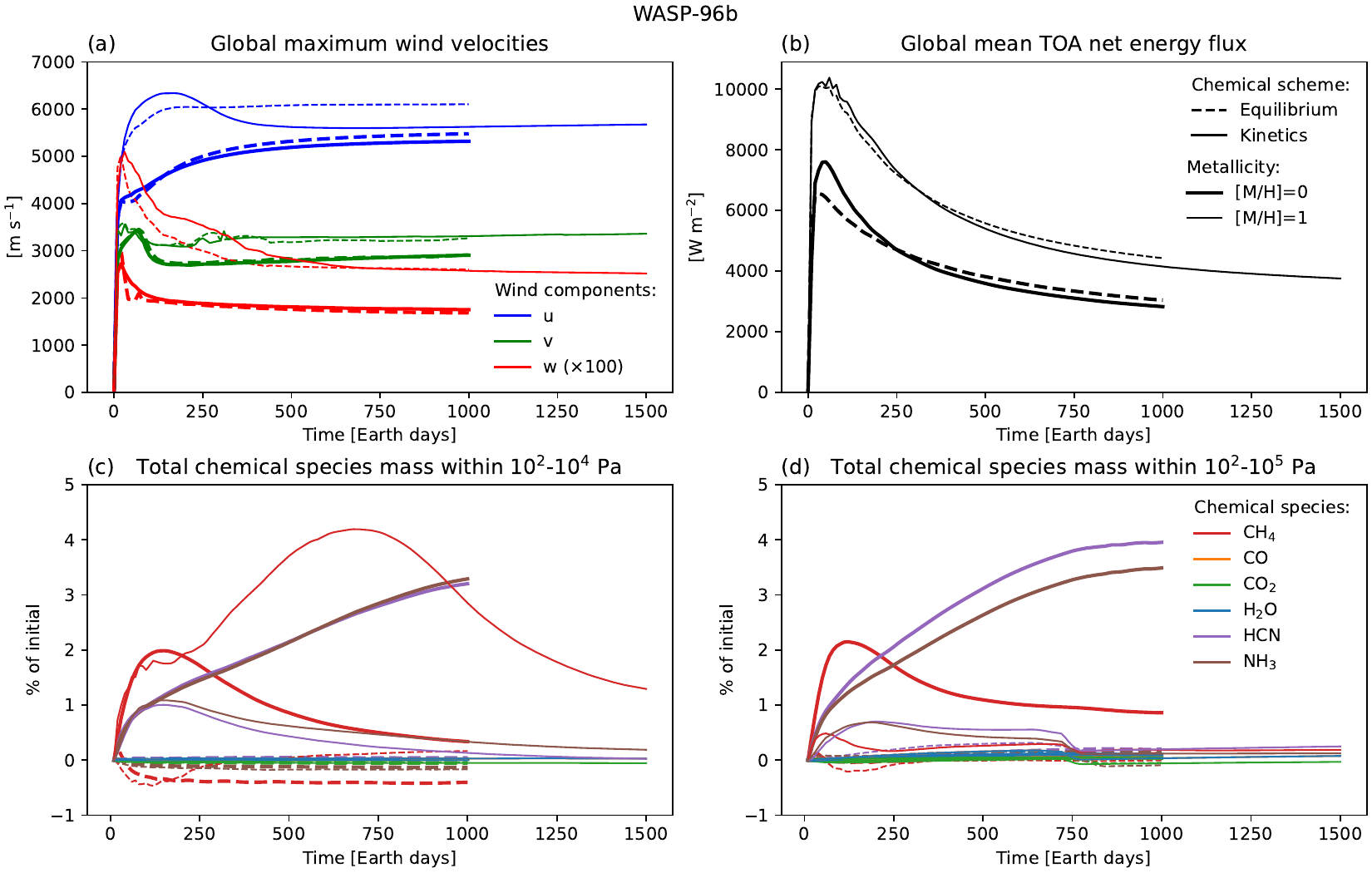}
 \caption{Steady state diagnostics from the WASP-96b equilibrium and kinetics simulations assuming $[M/H]=0$ or $[M/H]=1$. All panels share the legend of panel (b), and panels (c) and (d) share the legend of panel (d). Simulations approached the dynamical (a), radiative (b) and chemical (c)-(d) pseudo-steady state. The kinetics simulation with $[M/H]=1$ was run for longer (1500 Earth days) to better stabilise the total \ce{CH4} mass within \num{e2}-\num{e4} \si{\pascal}.}
\label{fig:wasp96b_steady_state}
\end{figure*}

\begin{figure*}
 \includegraphics[scale=0.5]{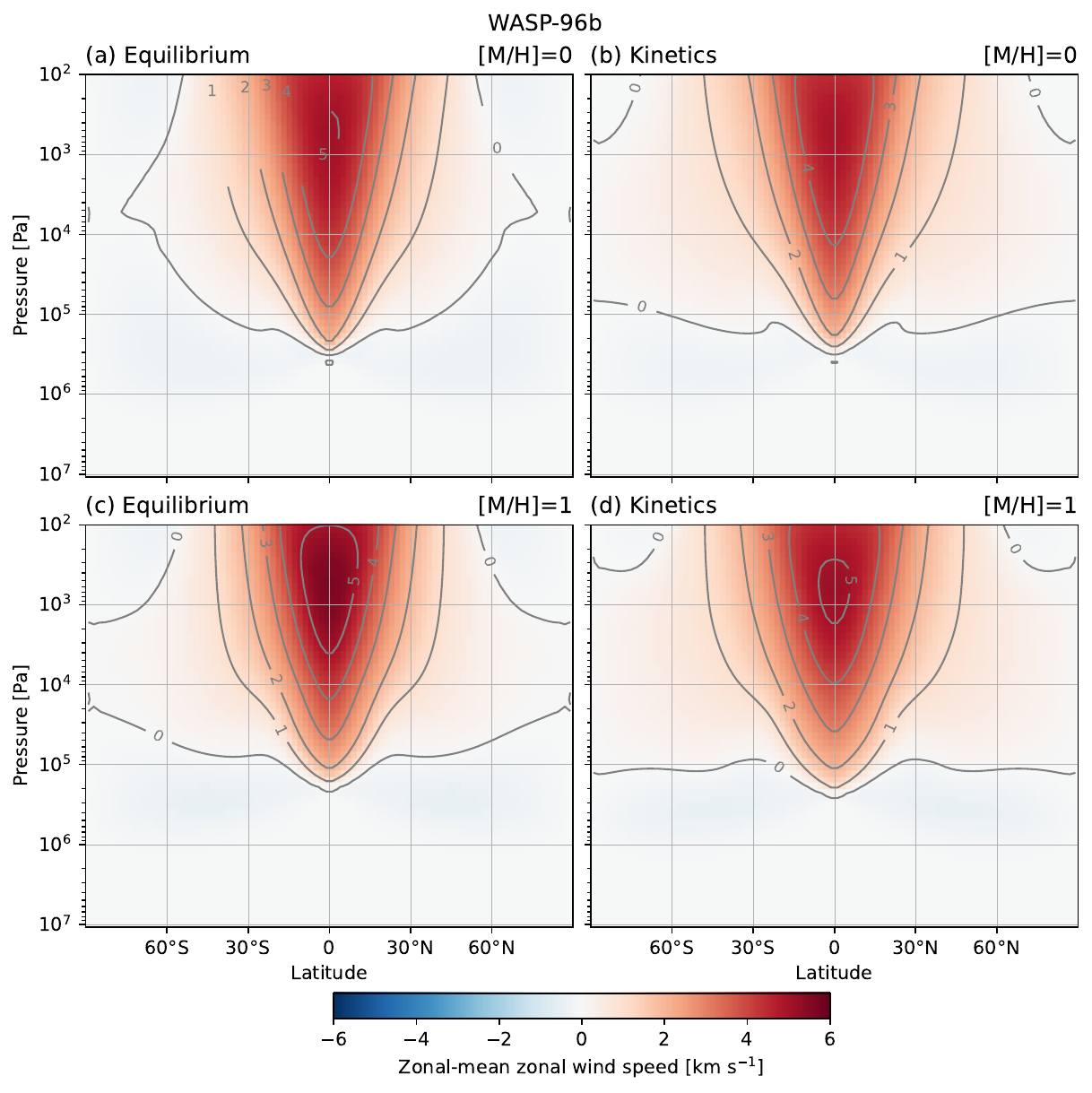}
 \caption{Zonal-mean zonal wind speeds from the WASP-96b equilibrium and kinetics simulations assuming $[M/H]=0$ or $[M/H]=1$. Simulations with a higher metallicity predict faster zonal-mean zonal wind speeds, but the general jet structure is similar in all simulations.}
 \label{fig:wasp96b_pres_u_znl_mean}
\end{figure*}

\begin{figure*}
    \includegraphics[scale=0.51]{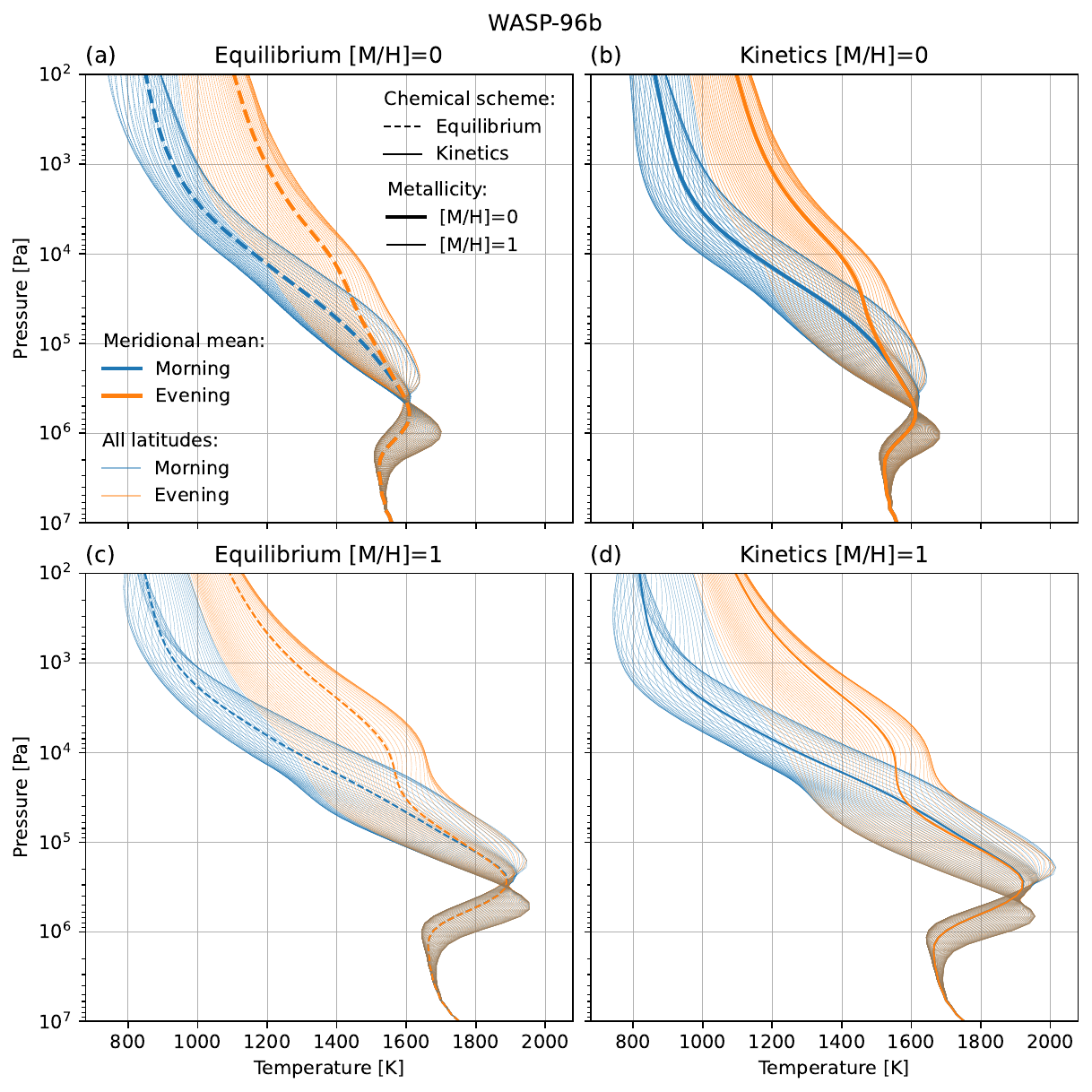}
    \caption{Pressure-temperature vertical profiles at the morning and evening limb from the WASP-96b equilibrium and kinetics simulations assuming $[M/H]=0$ or $[M/H]=1$.}
    \label{fig:wasp96b_vp_pres_temp_limbs}
\end{figure*}

\begin{figure*}
 \includegraphics[scale=0.58]{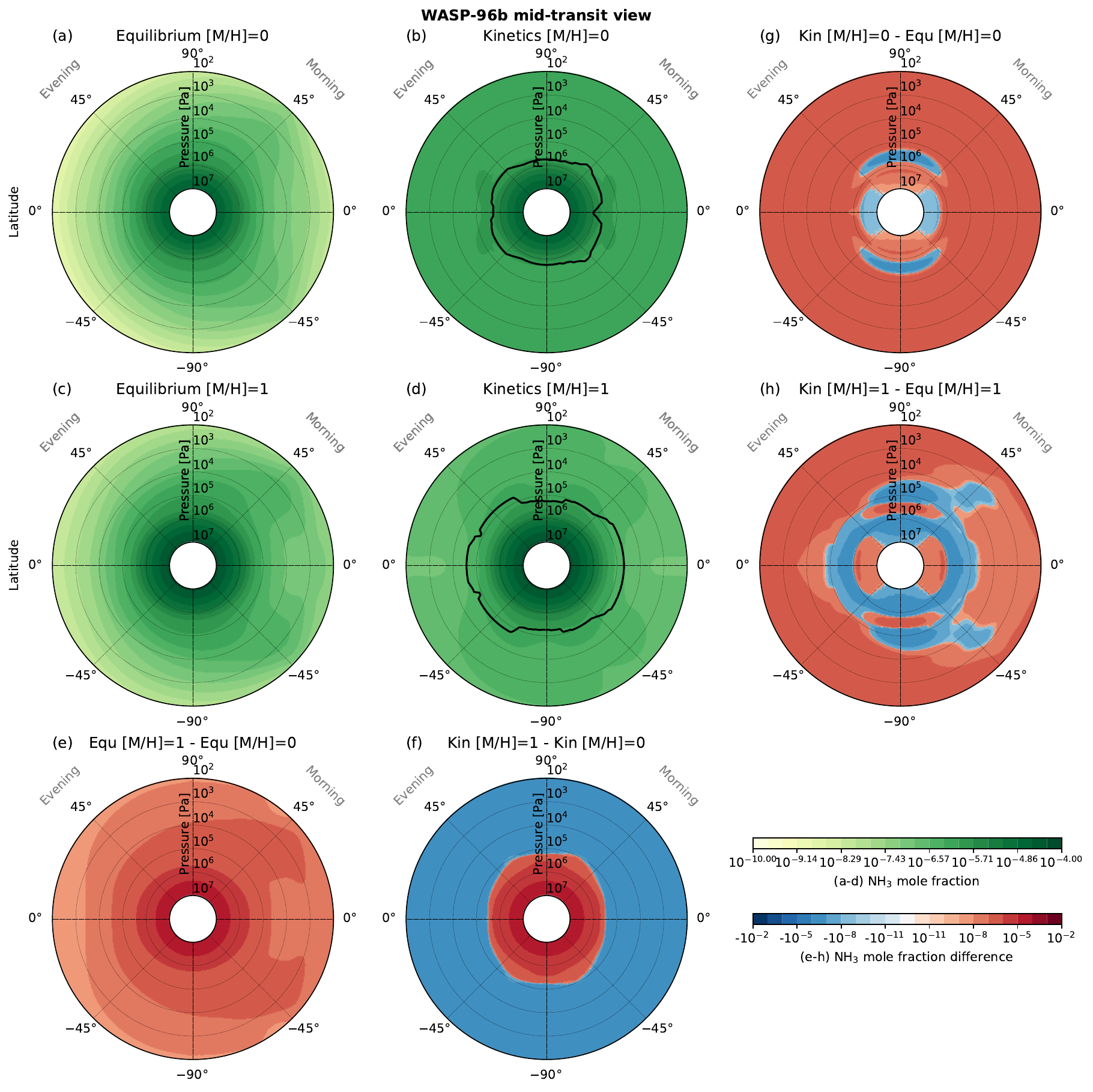}
 \caption{As in Fig.~\ref{fig:wasp96b_pres_ch4_in_transit} but for \ce{NH3}.}
\label{fig:wasp96b_pres_nh3_in_transit}
\end{figure*}

\begin{figure*}
 \includegraphics[scale=0.58]{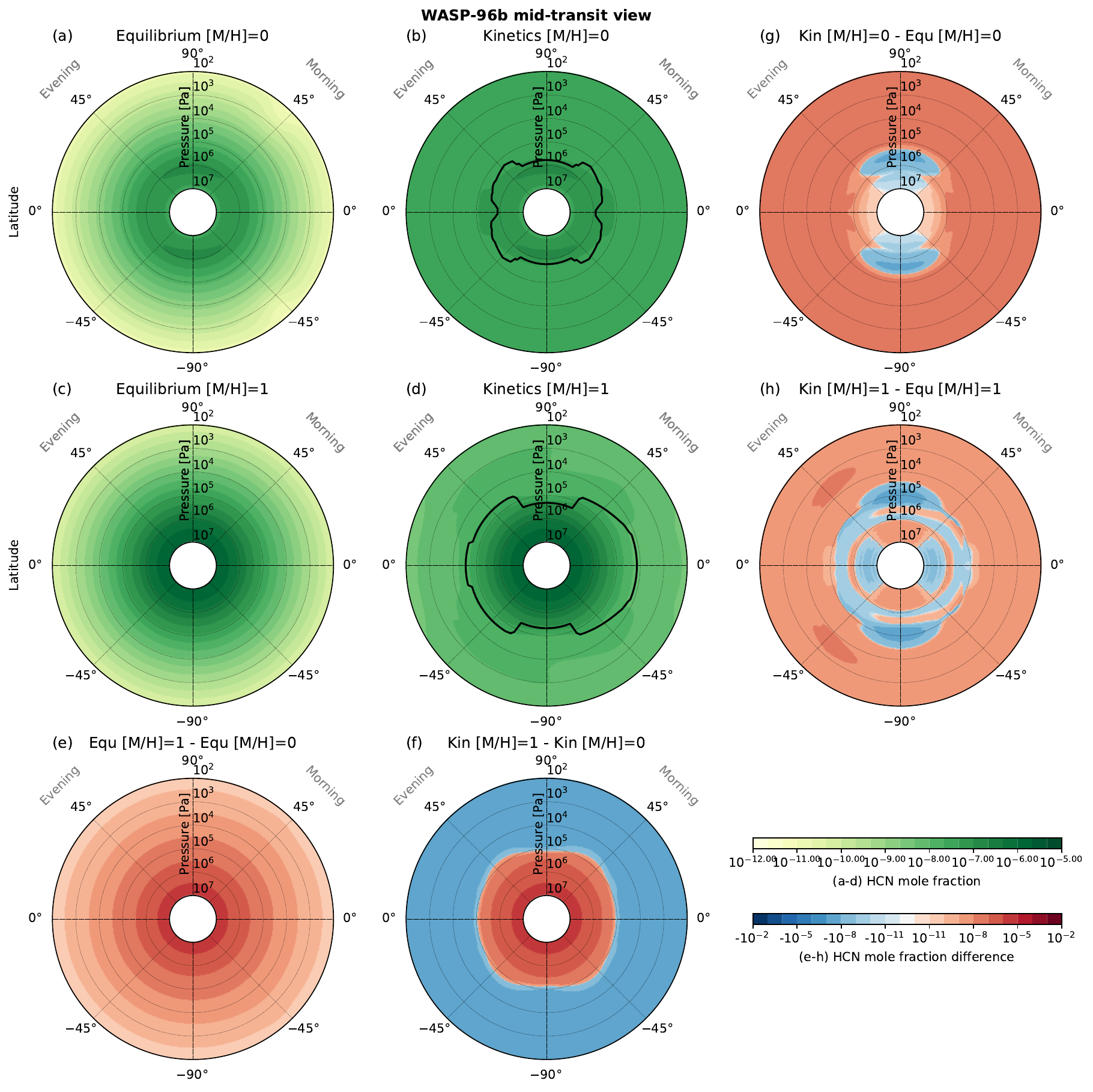}
 \caption{As in Fig.~\ref{fig:wasp96b_pres_ch4_in_transit} but for \ce{HCN}.}
\label{fig:wasp96b_pres_hcn_in_transit}
\end{figure*}

\begin{figure*}
 \includegraphics[scale=0.58]{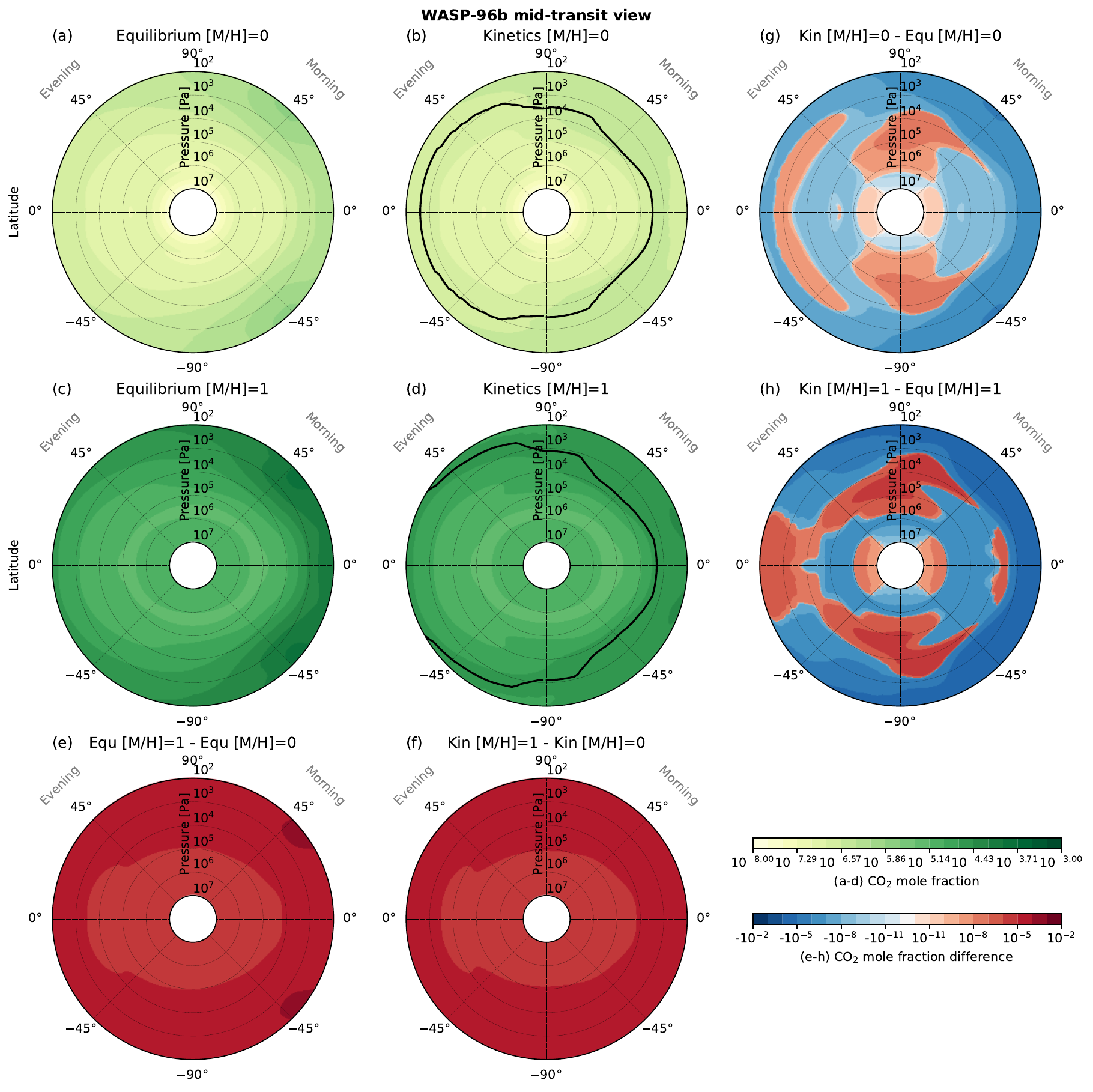}
 \caption{As in Fig.~\ref{fig:wasp96b_pres_ch4_in_transit} but for \ce{CO2}.}
\label{fig:wasp96b_pres_co2_in_transit}
\end{figure*}

\begin{figure*}
 \includegraphics[scale=0.58]{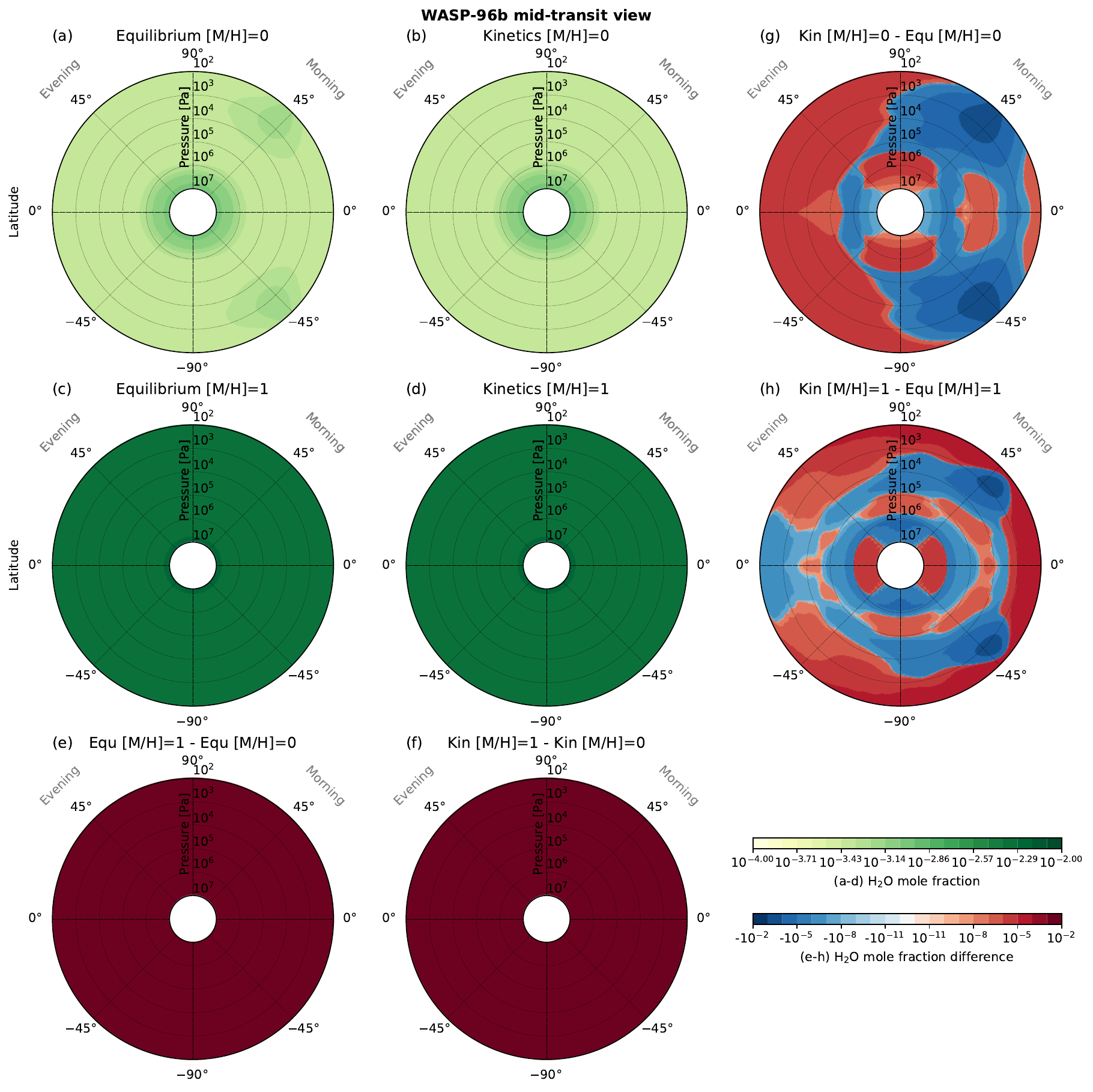}
 \caption{As in Fig.~\ref{fig:wasp96b_pres_ch4_in_transit} but for \ce{H2O}.}
\label{fig:wasp96b_pres_h2o_in_transit}
\end{figure*}

\begin{figure*}
 \includegraphics[scale=0.58]{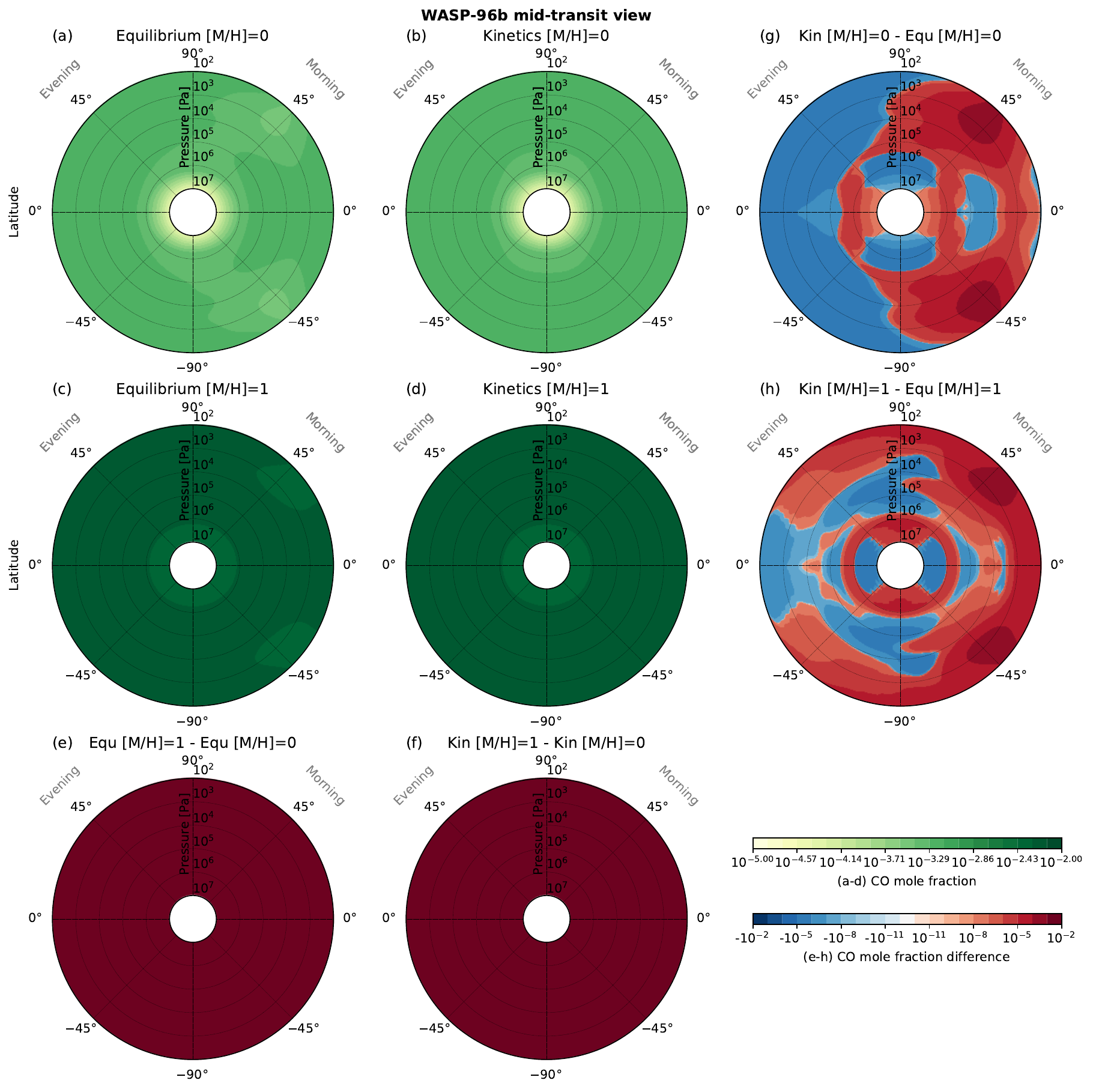}
 \caption{As in Fig.~\ref{fig:wasp96b_pres_ch4_in_transit} but for \ce{CO}.}
\label{fig:wasp96b_pres_co_in_transit}
\end{figure*}

\begin{figure*}
    \includegraphics[scale=0.51]{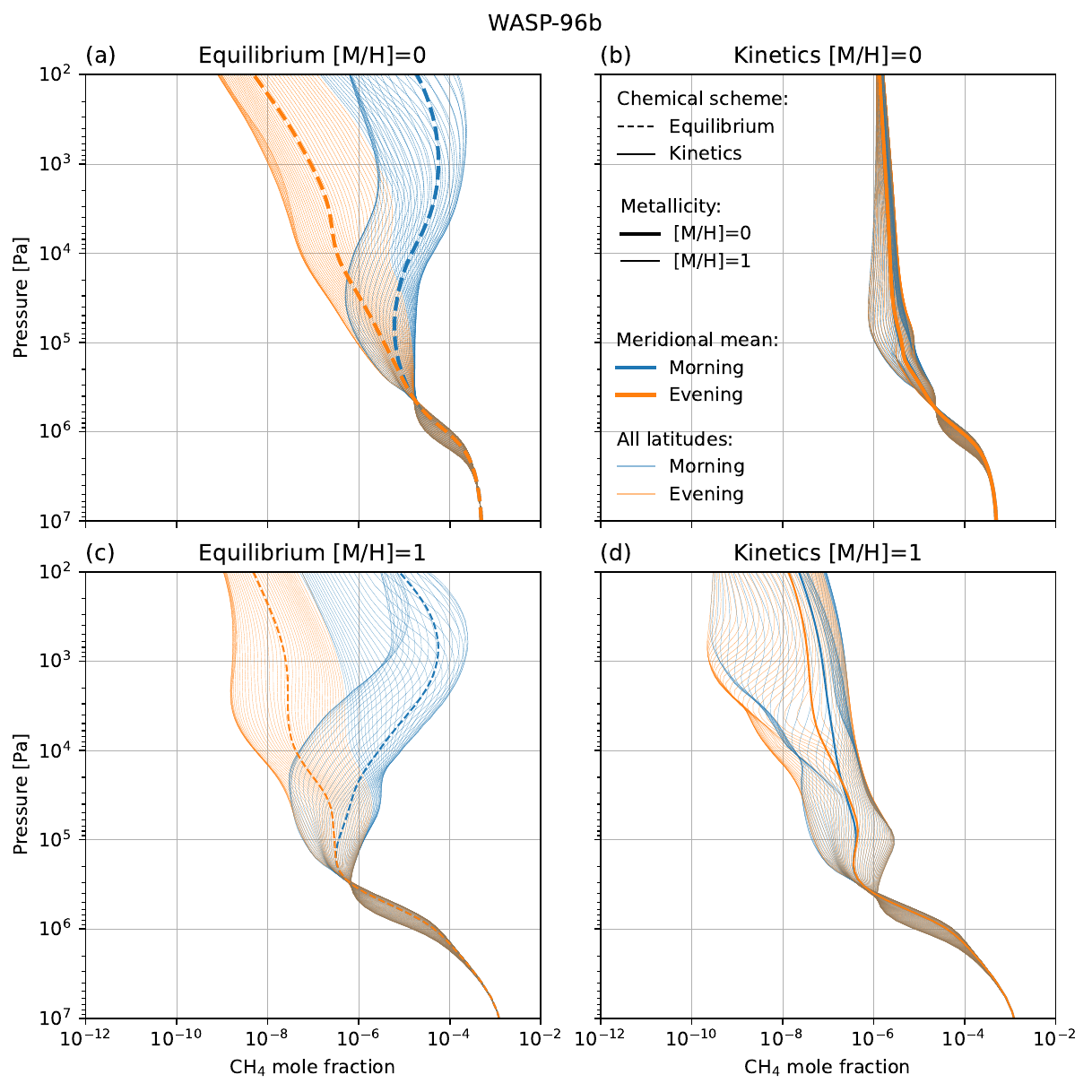}
    \caption{As in Fig.~\ref{fig:wasp96b_vp_pres_temp_limbs} but for \ce{CH4} mole fraction.}
    \label{fig:wasp96b_vp_pres_ch4_limbs}
\end{figure*}

\begin{figure*}
    \includegraphics[scale=0.51]{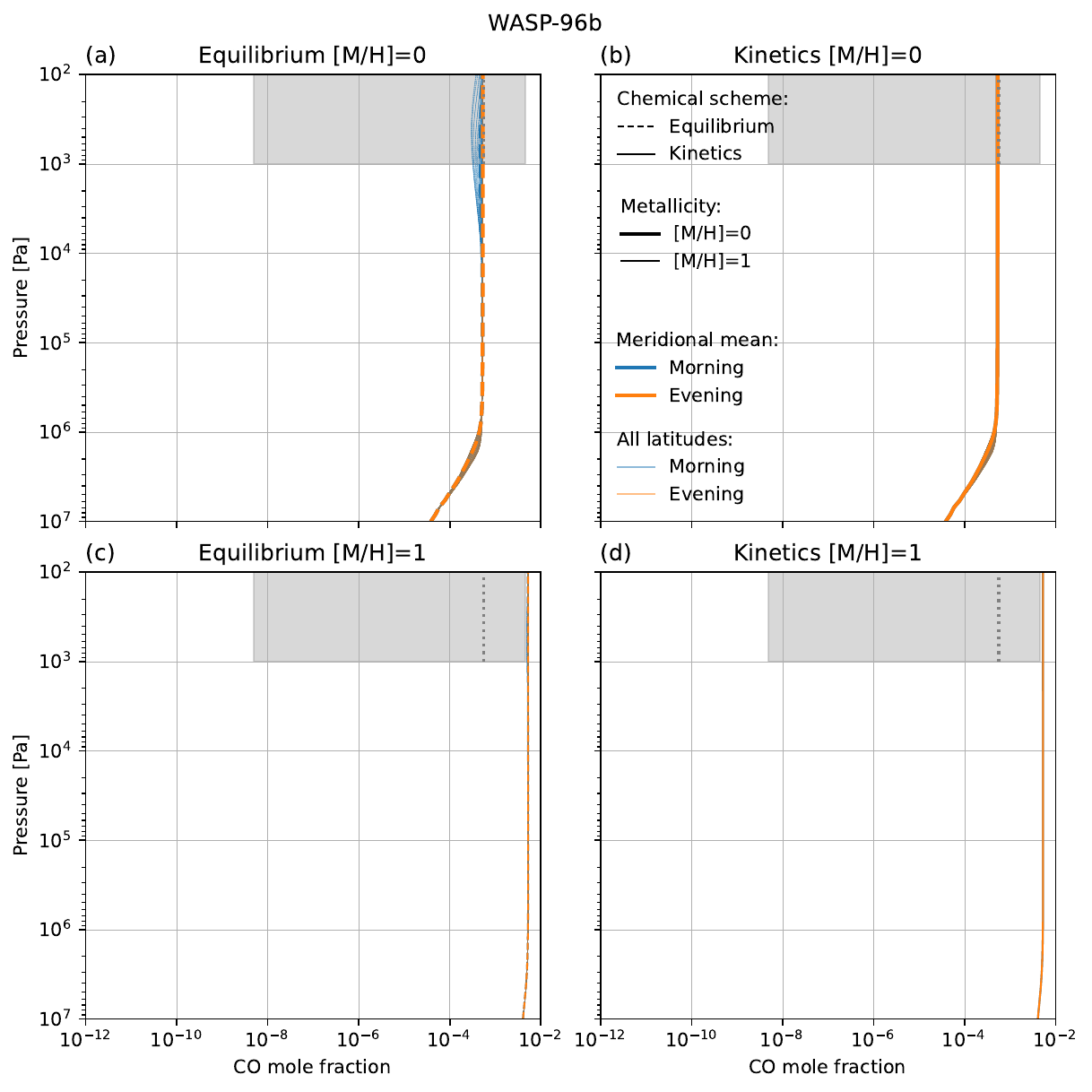}
    \caption{As in Fig.~\ref{fig:wasp96b_vp_pres_temp_limbs} but for \ce{CO} mole fraction. \ce{CO} abundance and its 1$\sigma$ uncertainty, $\log_{10}(\ce{CO})=-3.25${\raisebox{0.5ex}{\tiny$\substack{+0.91 \\ -5.06}$}}, recovered using the free retrieval code \textsc{Aurora} in \citet{Taylor2023} are shown as a grey dotted line and grey shading, respectively.}
    \label{fig:wasp96b_vp_pres_co_limbs}
\end{figure*}

\begin{figure*}
    \includegraphics[scale=0.51]{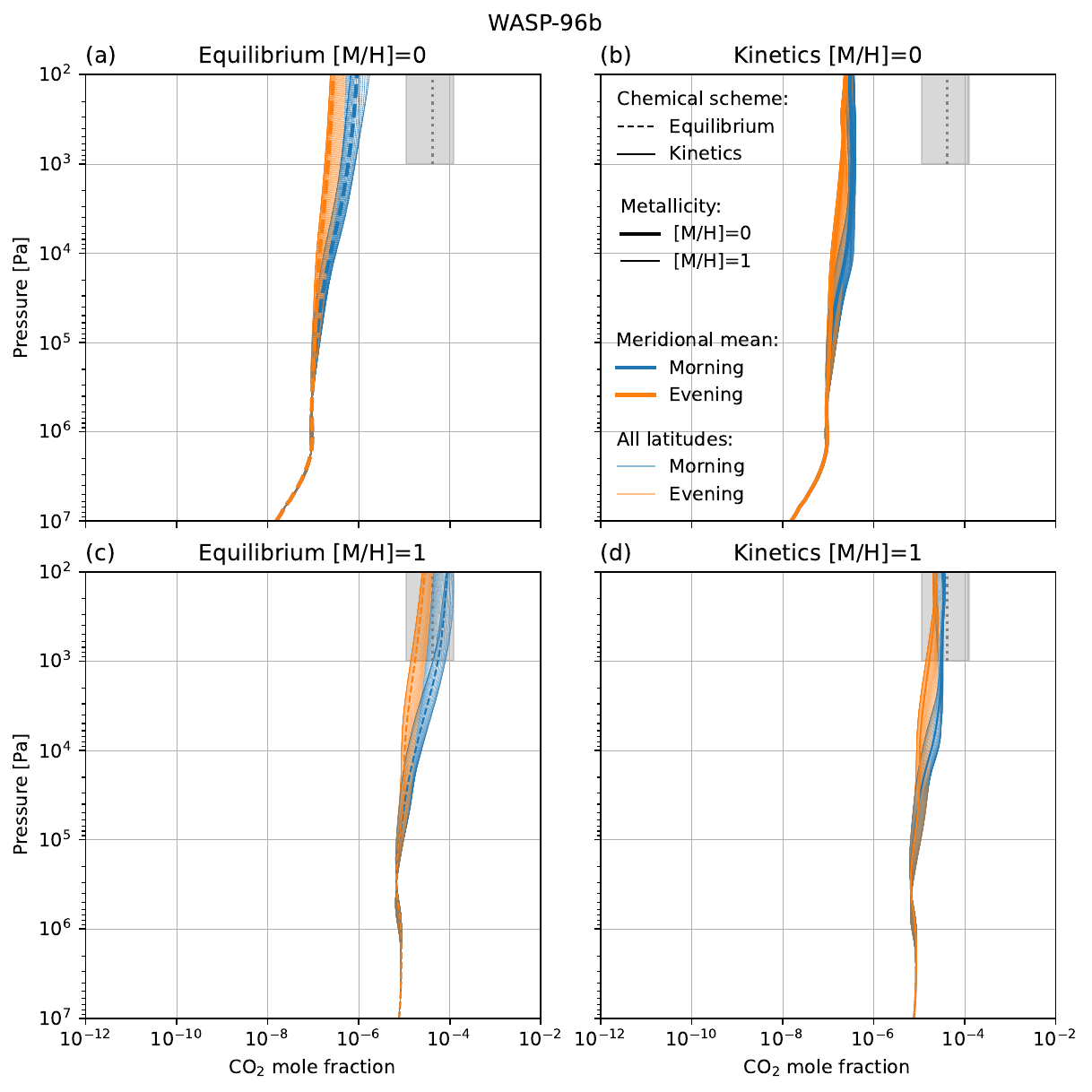}
    \caption{As in Fig.~\ref{fig:wasp96b_vp_pres_temp_limbs} but for \ce{CO2} mole fraction. \ce{CO2} abundance and its 1$\sigma$ uncertainty, $\log_{10}(\ce{CO2})=-4.38${\raisebox{0.5ex}{\tiny$\substack{+0.47 \\ -0.57}$}}, recovered using the free retrieval code \textsc{Aurora} in \citet{Taylor2023} are shown as a grey dotted line and grey shading, respectively.}
    \label{fig:wasp96b_vp_pres_co2_limbs}
\end{figure*}

\begin{figure*}
    \includegraphics[scale=0.51]{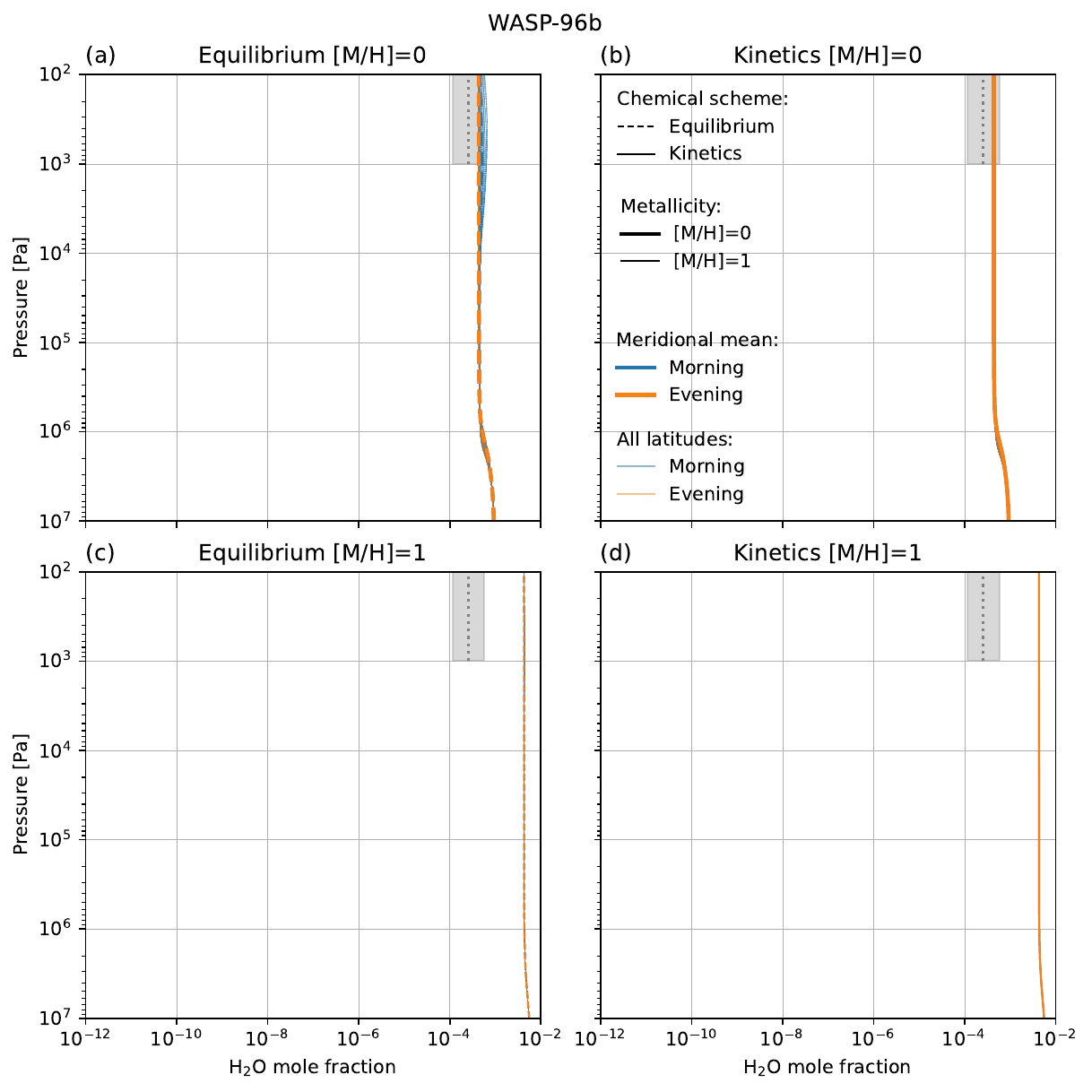}
    \caption{As in Fig.~\ref{fig:wasp96b_vp_pres_temp_limbs} but for \ce{H2O} mole fraction. \ce{H2O} abundance and its 1$\sigma$ uncertainty, $\log_{10}(\ce{H2O})=-3.59${\raisebox{0.5ex}{\tiny$\substack{+0.35 \\ -0.35}$}}, recovered using the free retrieval code \textsc{Aurora} in \citet{Taylor2023} are shown as a grey dotted line and grey shading, respectively.}
    \label{fig:wasp96b_vp_pres_h2o_limbs}
\end{figure*}

\begin{figure*}
    \includegraphics[scale=0.51]{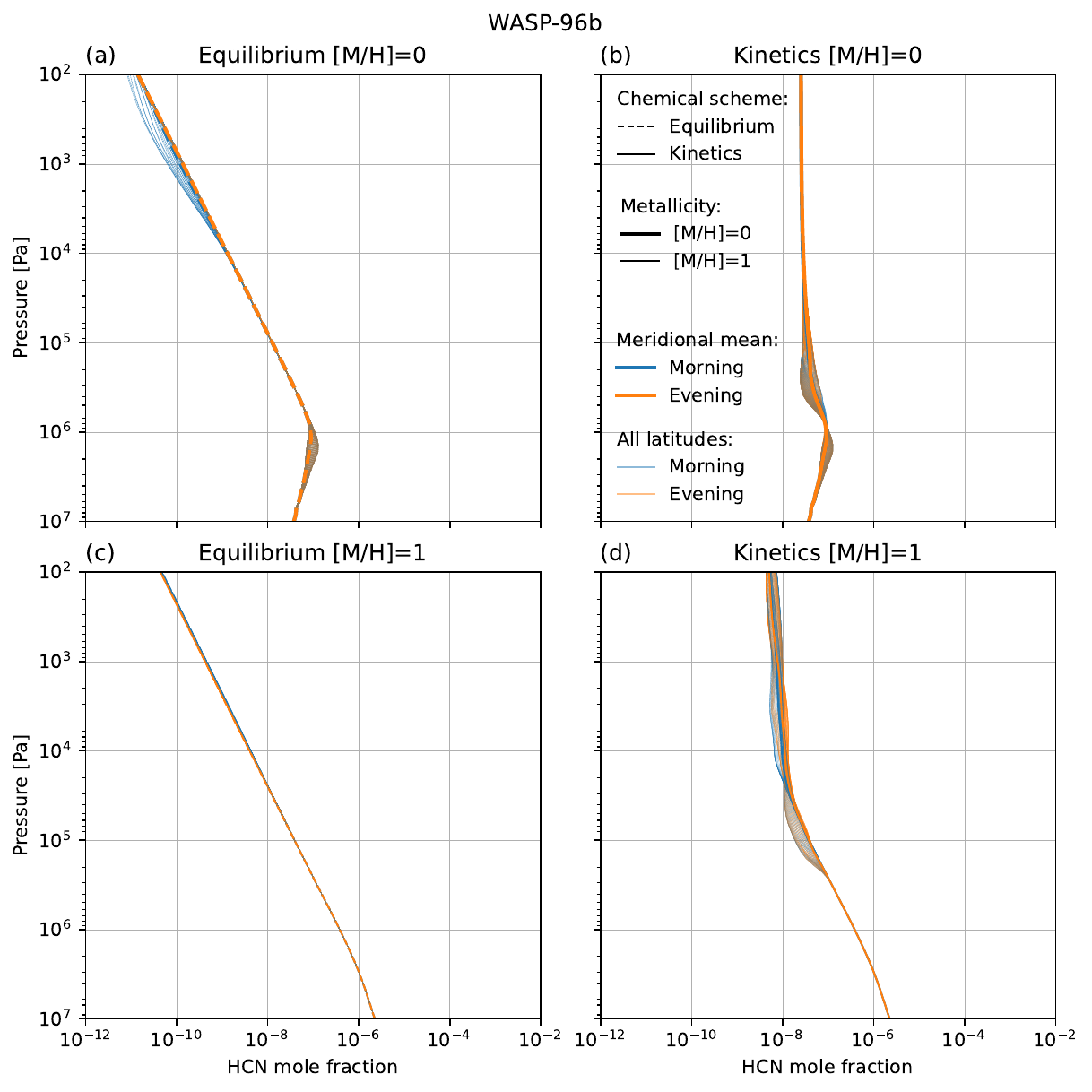}
    \caption{As in Fig.~\ref{fig:wasp96b_vp_pres_temp_limbs} but for \ce{HCN} mole fraction.}
    \label{fig:wasp96b_vp_pres_hcn_limbs}
\end{figure*}

\begin{figure*}
    \includegraphics[scale=0.51]{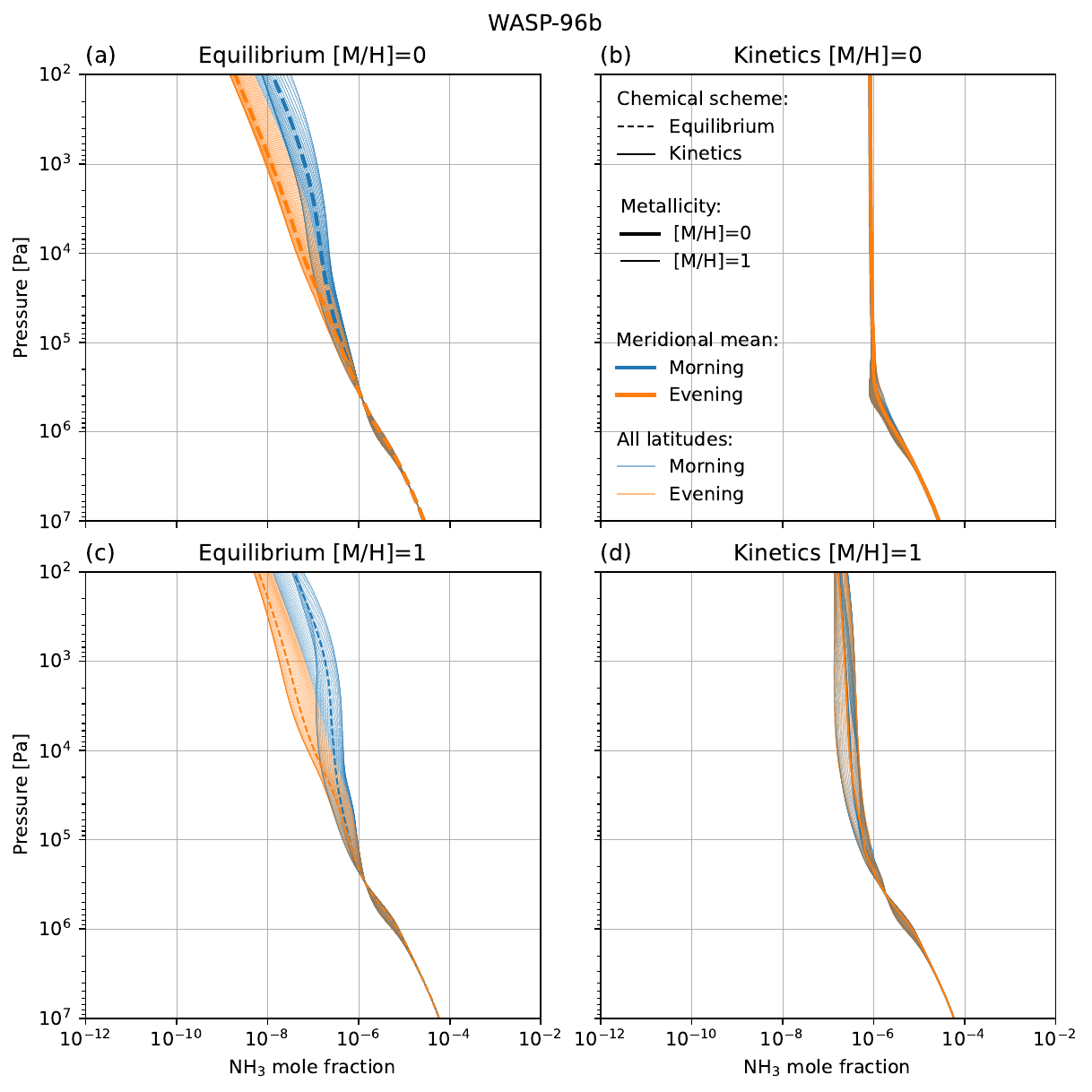}
    \caption{As in Fig.~\ref{fig:wasp96b_vp_pres_temp_limbs} but for \ce{NH3} mole fraction.}
    \label{fig:wasp96b_vp_pres_nh3_limbs}
\end{figure*}

\begin{figure*}
 \includegraphics[scale=0.55]{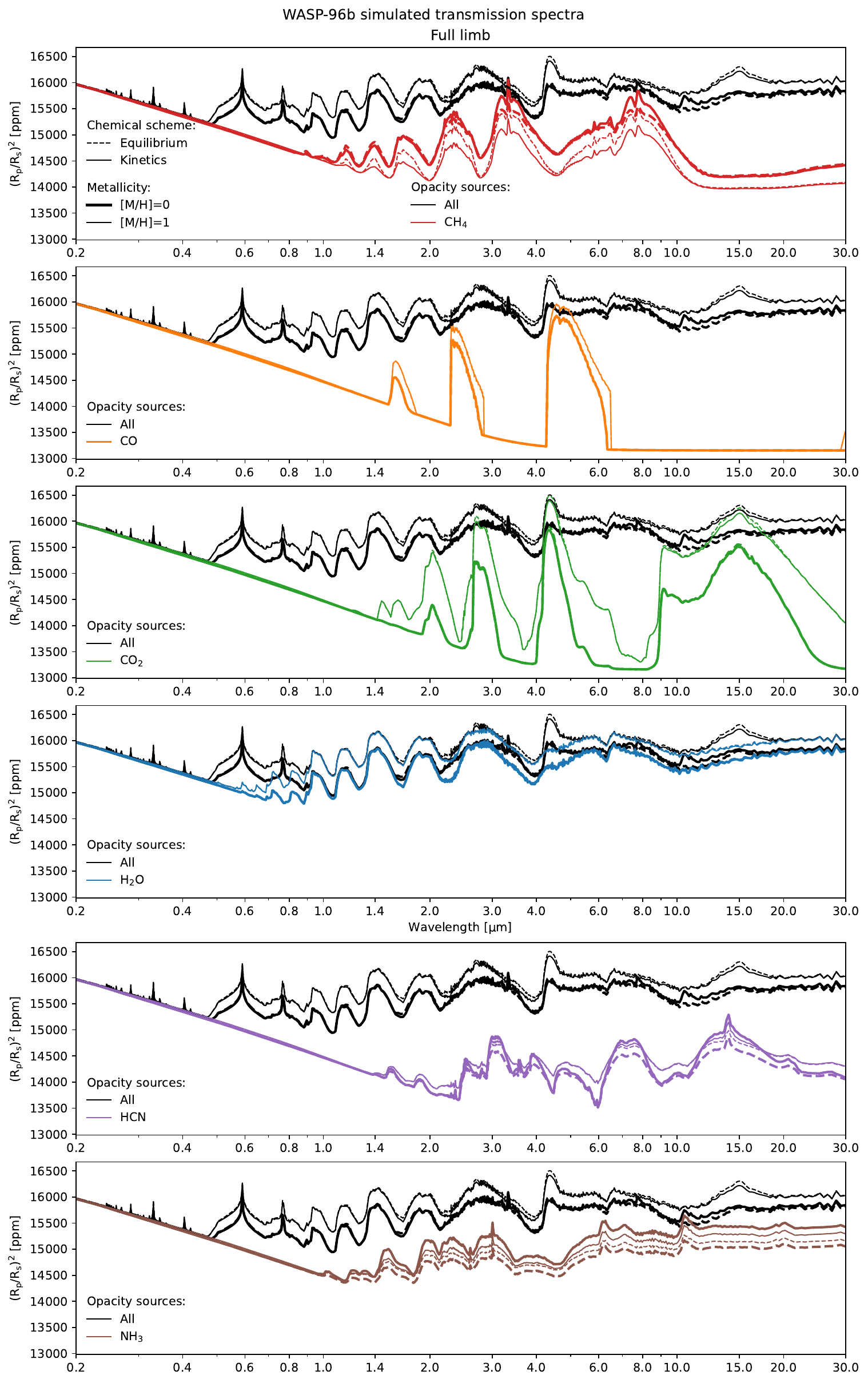}
 \caption{Contributions of \ce{CH4}, \ce{CO}, \ce{CO2}, \ce{H2O}, \ce{HCN} and \ce{NH3} to the full limb transmission spectra from the WASP-96b equilibrium and kinetics simulations assuming $[M/H]=0$ or $[M/H]=1$.}
\label{fig:wasp96b_transpec_wl02_30_combores_contribs_to_full}
\end{figure*}

\begin{figure*}
    \includegraphics[scale=0.62]{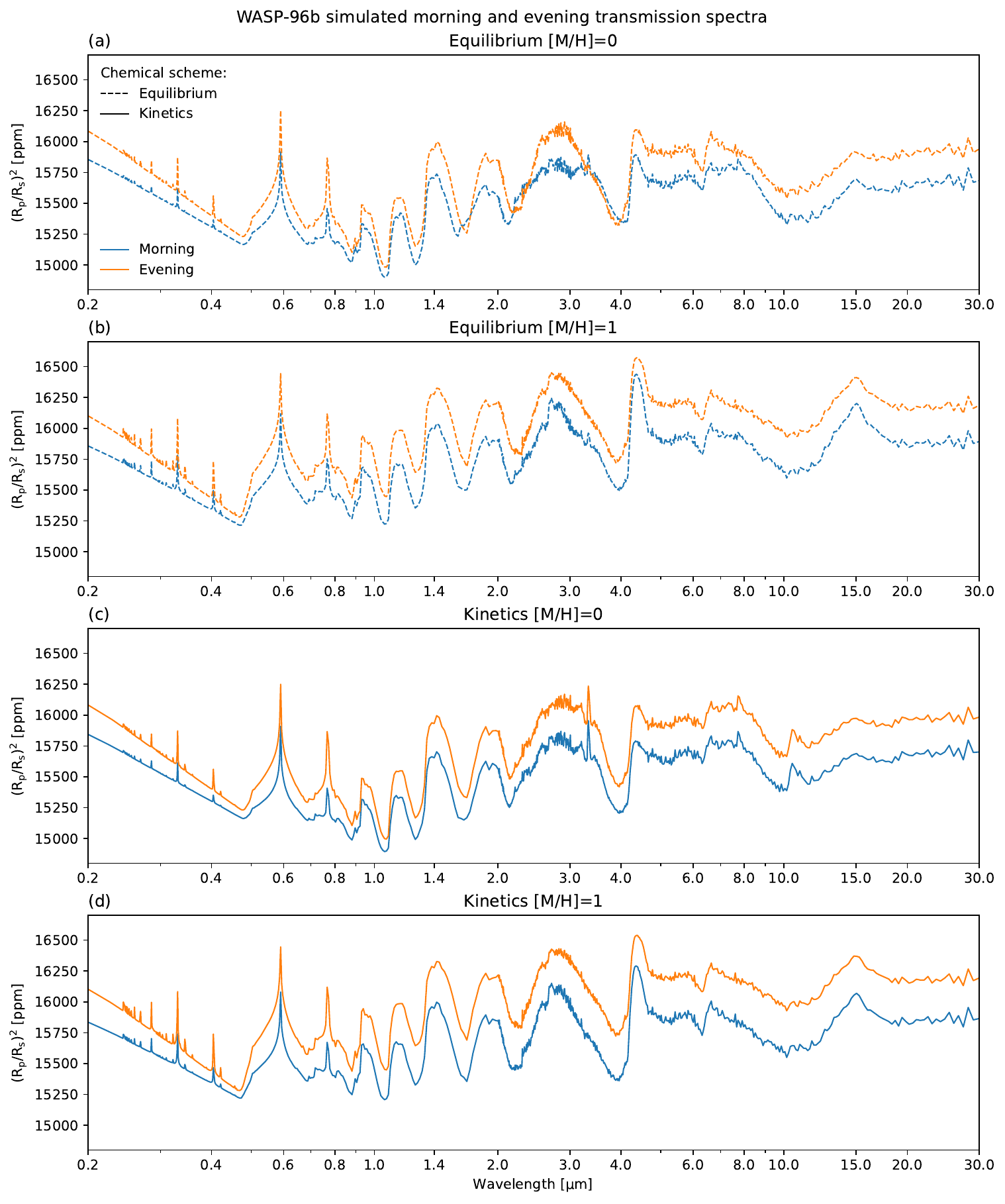}
    \caption{As in Fig.\ref{fig:wasp96b_transpec_wl02_30_combores_full_mor_eve}b but showing the morning and evening limb transmission spectra on separate panels.}
    \label{fig:wasp96b_transpec_wl02_30_combores_mor_eve}
\end{figure*}

Figures in this section provide additional information about our GCM simulations: Fig.~\ref{fig:wasp96b_conservation}-\ref{fig:wasp96b_steady_state} --- conservation and steady state diagnostics, Fig.~\ref{fig:wasp96b_pres_u_znl_mean} --- zonal-mean zonal wind speeds, Fig.~\ref{fig:wasp96b_vp_pres_temp_limbs} --- vertical pressure-temperature profiles in the terminator plane, Figs.~\ref{fig:wasp96b_pres_nh3_in_transit}-\ref{fig:wasp96b_pres_co_in_transit} --- the distribution of \ce{NH3}, \ce{HCN}, \ce{CO2}, \ce{H2O} and \ce{CO} in the terminator plane, Figs.~\ref{fig:wasp96b_vp_pres_ch4_limbs}-\ref{fig:wasp96b_vp_pres_nh3_limbs} --- vertical profiles of \ce{CH4}, \ce{CO}, \ce{CO2}, \ce{H2O}, \ce{HCN} and \ce{NH3}, Fig.~\ref{fig:wasp96b_transpec_wl02_30_combores_contribs_to_full} --- the contributions of \ce{CH4}, \ce{CO}, \ce{CO2}, \ce{H2O}, \ce{HCN} and \ce{NH3} to the full limb transmission spectra and Fig.~\ref{fig:wasp96b_transpec_wl02_30_combores_mor_eve} --- the morning and evening limb transmission spectra shown on separate panels.

\bsp	
\label{lastpage}
\end{document}